\documentclass[%
 reprint, pre,
superscriptaddress,
 amsmath,amssymb,
 aps,
floatfix,
]{revtex4-2}

\usepackage{graphicx}
\usepackage{dcolumn}
\usepackage{bm}
\usepackage{svg}
\usepackage{hyperref}
\usepackage{tikz}
\usepackage{tkz-euclide}
\usepackage{subcaption}

\usepackage{xcolor}

\begin{document}

\preprint{AIP/123-QED}

\title{Stochastic coupling of climate variables and ice volume over the Late Pleistocene glacial cycles} 
\author{Pijush Patra}
\affiliation{Nordita, KTH Royal Institute of Technology and Stockholm University, Stockholm 10691, Sweden} 
\author{Ludovico T. Giorgini}
\affiliation{Department of Mathematics, Massachusetts Institute of Technology, Cambridge, MA 02139, USA}
\author{J. S. Wettlaufer}
\homepage{Author to whom correspondence should be addressed: jw@fysik.su.se, john.wettlaufer@yale.edu}
\affiliation{Nordita, KTH Royal Institute of Technology and Stockholm University, Stockholm 10691, Sweden}
\affiliation{Departments of Mathematics, Earth \& Planetary Sciences, and Physics, Yale University, New Haven, Connecticut 06520-8109, USA}

\begin{abstract}
Understanding the interactions between ice sheets and global climate forcings over geological timescales is essential for projecting their future. Previous studies have highlighted the role of ice dynamics and climate interactions in establishing the 100,000-year glacial cycles, particularly regarding the growth of North American ice sheets. Researchers have reconstructed consistent time series for ice volume, temperature, and carbon dioxide by applying inverse forward modeling to benthic oxygen isotope records. In this paper, we model the stochastic behavior of paleoclimate time series to evaluate the coupling between climate variables during the Pleistocene glacial cycles. We quantify the behavior of these time series using multifractal time-weighted detrended fluctuation analysis, which differentiates between near-red-noise and white-noise behavior below and above the 100,000-year glacial cycle, respectively, in all records. This study builds upon the work of Keyes et al. [{\em Chaos} \textbf{33}, 093132 (2023)] by incorporating ice volume into a five-variable model that includes carbon dioxide, methane, nitrous oxide, and temperature, along with intervariable coupling terms to capture potential relationships among these variables. To test our model, we compute response functions for each pair of variables and compare them with empirical data, confirming our predictions regarding intervariable stability and coupling. This study provides a comprehensive overview of glacial-interglacial dynamics and highlights the role of cryosphere-climate feedbacks in shaping Earth's climate evolution.
\end{abstract}

\maketitle

{\bf The central challenge of understanding the pacing of the of Earth's ice ages during the last 800,000 years is to discern the relative importance of external forcing associated with orbital eccentricity, obliquity and precession, and internal dynamics associated with the processes of the atmosphere, biosphere, cryosphere and oceans.  This confluence of processes is encoded in proxies stored in deep sea sediments and ice cores, including greenhouse gases, temperature and ice volume.  Because the time series and co-variations among these variables are influenced by processes operating on multiple time scales, they evolve as stochastic processes.  Here we first analyze the structure of the noise in temperature, ice volume, $\text{CO}_2$, $\text{CH}_4$, and $\text{N}_2\text{O}$ using a method that provides colored-noise structure on multiple  timescales.  We then treat their co-variations as a coupled non-autonomous stochastic dynamical system.  Important results are  oscillatory dynamics associated with the couplings between ice volume, temperature and $\mathrm{CO}_2$, with a strong causal relationship between the latter driving the former two, which is the signature of the dominant role of atmospheric CO$_2$ as greenhouse forcing of glacial–interglacial transitions. Our approach also captures the observed lagged feedbacks of ice-volume on temperature and CO$_2$, consistent with the role of the massive ice sheets as a slow integrator of climate forcing.}

\section{Introduction} \label{Introduction}

\subsection{Background \& Motivation}

The glacial cycles of the Late Pleistocene (last $\sim 800$--$900$ kyr) are characterized by a periodicity of roughly $100$ kyr. Hays \textit{et al.} \cite{hays1976variations} linked this pacing of glacial cycles to Milankovitch theory, which attributes this periodicity to quasi-periodic alterations in Earth's orbital parameters: precession ($\sim 19$--$23$ kyr), obliquity ($\sim 41$ kyr), and eccentricity ($\sim 100$ kyr). These orbital variations modulate the seasonal and latitudinal distribution of solar insolation. However, the theory of orbitally driven glacial cycles encounters several complexities. One particularly compelling issue pertains to the observation that prior to $\sim 1$ Myr ago, fluctuations in the marine oxygen isotope ratio ($\delta^{18}\text{O}$) \cite{lisiecki2007plio} predominantly occurred at obliquity timescales . In contrast, these fluctuations transitioned to $\sim 100$ kyr timescales in subsequent epochs. This shift from 41-kyr to 100-kyr climatic cycles is widely recognized as the mid-Pleistocene transition (MPT), which occurred without any discernible change in insolation forcing. In recent years, numerous explanations have been proposed to explain MPT (see \citet{berends2021cause} for a recent review); however, the investigation of the physical mechanisms underpinning the 100-kyr cycle remains an active area of research.

Crucially, the pacing of glacial cycles by orbital eccentricity remains difficult to explain, as direct insolation forcing at the 100-kyr frequency is minimal \cite{wunsch2004quantitative}. Furthermore, formal hypothesis testing shows that eccentricity does not significantly modulate the timing of glacial terminations \cite{huybers2005obliquity}, although eccentricity variations are highly variable on multiple timescales, with chaotic behavior compromising precise solutions beyond 50 Myr \cite{Laskar2011}.  

\citet{Barker2025} have recently argued that, over the past 900 kyr, orbital forcing alone can explain both the occurrence and duration of all deglacial and interglacial periods, with the initiation of deglaciation associated with the combination of precession and obliquity, and glacial inception controlled solely by obliquity.  In contrast, \citet{Zhang2025} propose that the nonlinear interaction between eccentricity-modulated precession and the internal coupling between ice volume and the carbon cycle explain the 100-kyr cycle, and \citet{Hobart2023Precession100kyr} argue that precession, and thus Northern Hemisphere summer insolation intensity, is most predictive of onset of glacial termination.  Taken together, these findings suggest that it remains hotly debated whether orbital forcing alone can account for the pronounced 100-kyr cycle, thereby implying that internal climate dynamics and feedbacks must play a central role in shaping the glacial cycle.

In addition to those discussed above, several internal feedback mechanisms have been proposed to explain the emergence of the 100-kyr cycle, including enhanced North American ice-sheet growth and the subsequent merging of the ice sheets \cite{bintanja2008north}, spatiotemporal evolution of ice-sheet substratum \cite{clark1998origin}, carbon dioxide variability \cite{shackleton2000100}, oceanic carbon uptake, carbon rock weathering, soil nitrogen release, and dust feedbacks \cite{dean2018methane,williams2019carbon,prentice2012modelling}. However, the sequence, relative importance, and interactions of these feedbacks remain uncertain, limiting our understanding of the coupled dynamics among temperature, ice-sheet evolution, and greenhouse-gas forcing. Global ice volume, in particular, plays a central role in the climate system as both a slow-moving integrator of climate forcing and a strong feedback mechanism via albedo, topography, and ocean–carbon interactions. The Late Pleistocene glacial cycles, therefore, provide a valuable opportunity to explore linkages between components of the climate system, ice-sheet behavior, and Earth system sensitivity to greenhouse-gas forcing, as well as to assess the processes that may have underlain the pacing and amplitude of past glaciations.

While our aim here is not to propose a new mechanistic theory of glacial pacing, we seek to characterize the stochastic dynamics, noise structure, and causal relationships among key paleoclimate variables associated with Late Pleistocene glaciations. To achieve this, we focus on developing models that can reproduce multiscale stochastic dynamics while elucidating the causal interactions among variables. Conventional statistical techniques, such as covariance analysis, can quantify the strength of associations between variables, but they cannot determine the direction of influence or whether one process exerts stabilizing or destabilizing effects on another. Directionality, in principle, can be inferred through the application of a generalized Fluctuation–Dissipation Relation \cite{baldovin2022extracting,giorgini_response_theory,giorgini2025predicting,cooper2011climate,baldovin2020understanding,ghil2020physics,majda_climate_response,MajdaBook, lucarini2017predicting}, which can provide estimates of causal links among variables, but cannot differentiate between stabilizing and destabilizing interactions. Global climate models represent another approach, explicitly simulating atmosphere–ocean interactions by numerically integrating the governing conservation equations and incorporating external forcing, and parameterizations of unresolved small-scale processes \cite{hansen1983efficient}. Despite their physical rigor, these models frequently encounter challenges in accurately reproducing the full spectrum of variability, small-scale structural features, and long-duration time series of the variables involved. These limitations are partly attributable to uncertainties in subgrid-scale parameterizations and inter-model discrepancies, as well as the substantial computational cost associated with simulations that run over long time periods \cite{stone1990limitations,lopez2020generalized,alizadeh2022advances}.

Recent paleoclimate studies have examined causal relationships among proxy time series using a variety of methodological frameworks. These include approaches such as prediction skill through convergent cross-mapping \cite{van2015causal}, lead-lag estimations that quantify temporal lags between atmospheric carbon dioxide concentrations and temperature during glacial transitions \cite{fischer1999ice}, information-theoretic measures that estimate directional information flow among variables \cite{stips2016causal}, and parametric statistical models like multivariate autoregressive modeling \cite{kaufmann2016testing}. Additionally, linear inverse methods grounded in the Liang–Kleeman formalism have been developed to quantify directional information transfer and causal influence in dynamical systems by explicitly linking time series evolution to underlying governing equations \cite{lien2025linear}. Multifractal methods have also been employed to investigate scale-dependent dynamics in paleoclimate records. For example, \citet{shao2016contrasting} applied a multifractal approach to Antarctic and Greenland ice-core data, demonstrating that the Holocene climate is predominantly monofractal, while the glacial climate exhibits multifractal properties, and concluded that the glacial climate has a longer persistence time and stronger nonlinearities. Using a similar multifractal method, we have recently assessed the scaling dynamics of temperature and three greenhouse gas time series extracted from ice cores drilled at Dome C in Antarctica by the European Project for Ice Coring in Antarctica (EPICA) project \cite{keyes2023stochastic}. In the current study, we extend the analysis by including ice volume time series along with temperature, carbon dioxide ($\text{CO}_2$), methane ($\text{CH}_4$), and nitrous oxide ($\text{N}_2\text{O}$). \textcolor{black}{Rather than merely an increase in model dimensionality, this extension is dynamically important, enabling us to assess which interaction patterns identified in prior studies remain valid and which are altered when ice-sheet dynamics are included.}

Our approach employs a stochastic data analysis and modeling framework grounded in the characterization of colored noise and the theory of non-autonomous stochastic dynamical systems. Similar to other stochastic dynamical systems methodologies in climate science \cite{majda1999models,ghil2020physics,giorgini2022non,giorgini2025reduced, giorgini2025score, falasca2026causally, franzke2015stochastic}, this methodology enables us to characterize random variability inherent in climate processes that purely deterministic models fail to capture.

We begin by quantifying the noise characteristics of the proxy time series using multifractal time-weighted detrended fluctuation analysis (MFTWDFA) \cite{zhou2010multifractal}. This method enables us to identify the color of the noise and understand how these characteristics differ across varying timescales. Such an analysis is particularly pertinent for climate records exhibiting considerable stochastic variability and distinct timescale separation, as it provides insights into the differing underlying dynamics that define shorter and longer temporal regimes.

We employ Ornstein–Uhlenbeck processes to model paleoclimate time series, formulating them as periodic, non-autonomous Langevin equations that explicitly account for both the deterministic behavior and stochastic variability of the data. To assess the performance of our model, we compute response functions for each pairwise combination of variables after applying the model to time series of temperature, ice volume (IV), $\text{CO}_2$, $\text{CH}_4$, and $\text{N}_2\text{O}$.

Finally, we use the estimated model coefficients to generate synthetic realizations of the time series and assess how well they reproduce original records using multiple statistical performance measures. We subsequently evaluate response functions and elucidate their physical implications for the interactions among these paleoclimate variables.

The remainder of this paper is organized as follows. In Sec. II, we apply MFTWDFA to characterize the noise structure of the paleoclimate records. We formulate and implement the Ornstein–Uhlenbeck models in Sec. III and evaluate the model fidelity and investigate causal relationships among the variables. Finally, in Sec. IV, we conclude with a discussion of the implications of our findings for the evolution of Earth’s climate over the past 800,000 years.

\begin{figure*}
\centering
\includegraphics[width=1.0\textwidth]{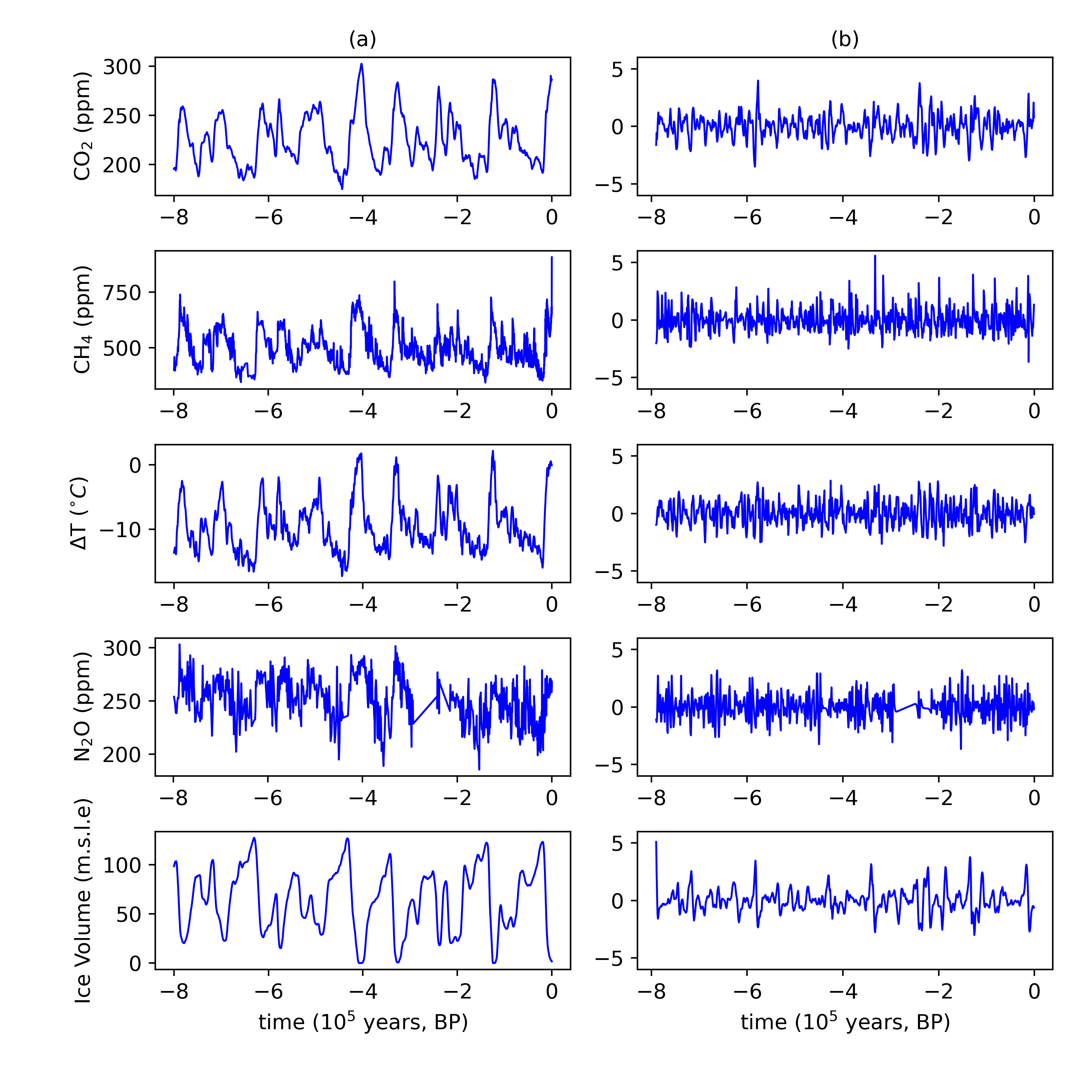}
\caption{Carbon dioxide, methane, temperature, nitrous oxide, and ice volume time series. (a) Original time series, (b) normalized fluctuations time series relative to the slowly varying mean. Here BP denotes before present and m.l.s.e denotes meters sea level equivalent.}
\label{fig:data}
\end{figure*}

\begin{figure*}
\centering
\includegraphics[width=1.0\textwidth]{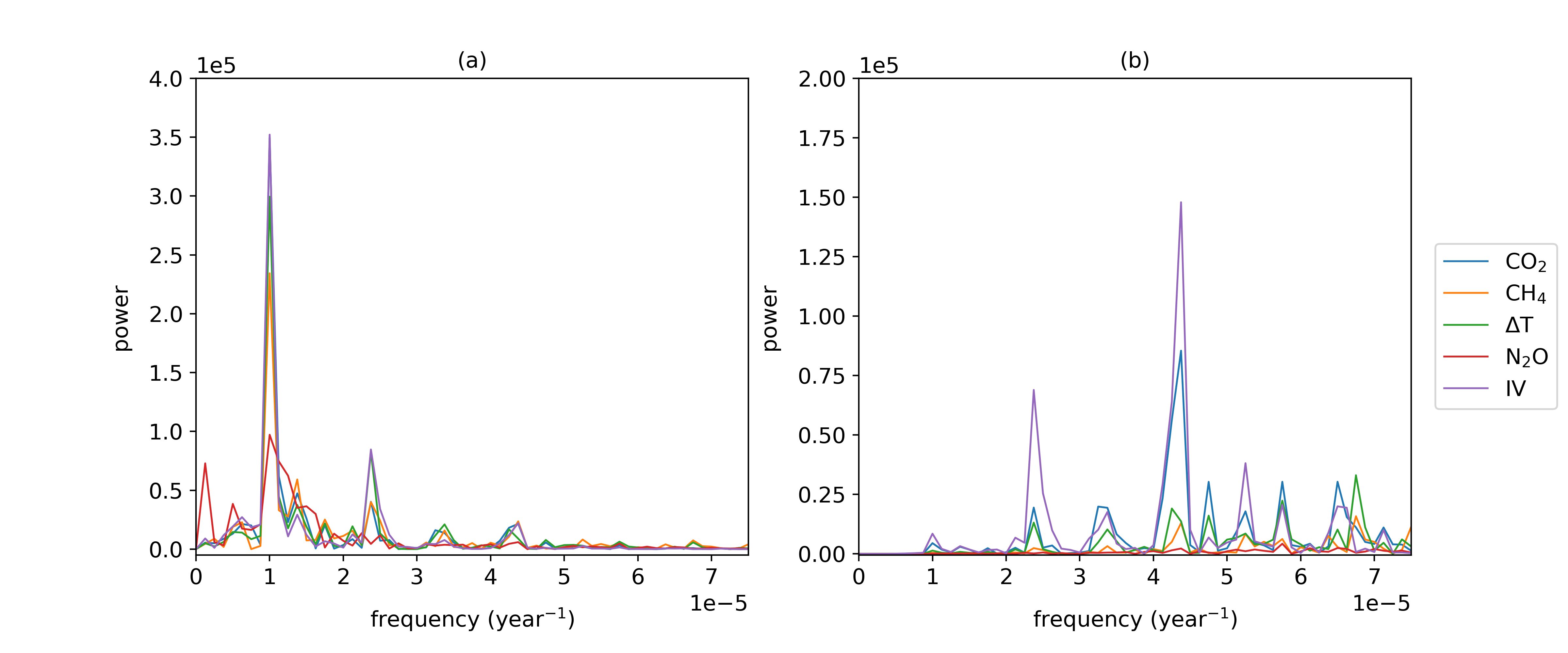}
\caption{Frequency spectra of (a) original time series and (b) normalized fluctuations time series relative to the slowly varying mean.}
\label{fig:spectra}
\end{figure*}

\subsection{The dataset} \label{The_dataset}

\subsubsection{Background}

Ice cores and marine sediment cores are the two primary sources of information for the Late Pleistocene glacial cycles. Ice cores provide two key categories of paleoclimate information: the heavy water isotopes ($\delta$D and $\delta^{18}\text{O}$), which serve as proxies for temperature reconstructions, and firn air trapped in bubbles and clathrates in the ice, which allow direct measurement of past atmospheric concentrations of the three major greenhouse gases ($\text{CO}_2$, $\text{CH}_4$, and $\text{N}_2\text{O}$) \cite{jouzel2007orbital}. Currently, the EPICA Dome C ice core provides a climate record that extends back to 800 kyr before present (BP). In line with the study by Keyes \textit{et al.} \cite{keyes2023stochastic}, we use the datasets for $\text{CH}_4$ and $\text{N}_2\text{O}$ following the EDC3 chronology \cite{parrenin2007edc3}.

Global ice volume is conventionally reconstructed from marine oxygen isotope ($\delta^{18}\text{O}$) records, which are influenced by both the deep-sea temperature and the size of ice sheets. In recent years, the application of inverse forward modeling has emerged as a powerful technique to disentangle the contributions of ice volume and deep-ocean temperature from the benthic $\delta^{18}\text{O}$ signal \cite{bintanja2005modelled,berends2021reconstructing}. This method involves a forward model that links $\delta^{18}\text{O}$ to temperature and ice volume, combined with an inverse step that iteratively adjusts these climate variables so that the modeled $\delta^{18}\text{O}$ matches the observed record. Using the LR04 benthic $\delta^{18}\text{O}$ stack \cite{lisiecki2005pliocene}, which is compiled from 57 globally distributed sediment cores, researchers have reconstructed mutually consistent time series of ice volume, temperature, and CO$_2$ \cite{bintanja2008north,berends2021reconstructing}. Berends \textit{et al.} \cite{berends2021reconstructing} illustrated that the reconstructed CO$_2$ time series agrees well with observations from the EPICA Dome C ice core record. In this study, we analyze the last 800 kyr of these reconstructed time series.

As shown in Figure \ref{fig:data}a, the characteristic 100 kyr periodicity of glacial cycles is clearly evident in the time series. In addition to this dominant periodicity, however, the datasets exhibit complex and intrinsically noisy structure across multiple timescales. To investigate this structure further, we first compute the frequency spectrum of all five time series, as shown in Figure \ref{fig:spectra}a. The spectra reveal a pronounced peak near $10^{-5}$ $\text{year}^{-1}$, corresponding to the 100-kyr periodicity, which aligns with the eccentricity cycle. It is pertinent to note that lower-frequency peaks are not reliable due to the lack of sufficient data. As mentioned above, this prominent 100-kyr glacial cycle has been the subject of extensive debate and investigation in the paleoclimate community.

The frequency spectra (Figure \ref{fig:spectra}a) also show additional significant peaks at approximately $2.5 \times 10^{-5}$ $\text{year}^{-1}$ and $4.25 \times 10^{-5}$ $\text{year}^{-1}$, which correspond to the $41$ kyr obliquity cycle and $23.5$ kyr net precession (resulting from the combination of axial and apsidal precession) cycle, respectively. These peaks are typically interpreted as manifestations of external astronomical forcing present in all sufficiently long time series, leading to strong correlations among them that complicate unraveling their causal relationships \cite{hays1976variations}. To mitigate the influence of this external forcing, we applied a high-pass filter that involved subtracting a running average from each time series (see Sec. \ref{Data_preparation} for additional details). The resulting spectra, shown in Figure \ref{fig:spectra}b, demonstrate that while the $10^{-5}$ $\text{year}^{-1}$ peak is substantially diminished, the peaks corresponding to the obliquity and net precession are enhanced. \textcolor{black}{Since each filtered fluctuation time series is subsequently normalized to unit variance, the total integrated spectral power is identical for all variables, in accordance with Parseval's theorem. Consequently, the differences observed in Figure \ref{fig:spectra}b reflect variations in the distribution of spectral power across frequencies, rather than differences in the total spectral power.} Thus we have filtered out, rather than entirely eliminated, this time-varying external forcing, thereby retaining an associated footprint in the time series that requires careful consideration. These observations motivate the development of a non-autonomous stochastic treatment of the filtered data to investigate the inherent causal relationships among the time series data.

\subsubsection{Data preparation} \label{Data_preparation}

To apply the stochastic modeling approach, we first interpolate each time series to an evenly spaced temporal grid. We choose a target resolution that closely matches the coarsest dataset—nitrous oxide, comprising 912 data points—and split the full 800 kyr time domain into 34 equal segments, yielding 25 points per segment and an average time spacing of approximately 929 years. Although this uniform spacing implies that multiple points may be interpolated into certain time gaps in the original series, we have confirmed that this occurrence is relatively rare. Most time gaps in the original dataset closely correspond to the interpolation gap, with the primary exception being a significant gap observed in the nitrous oxide time series, a limitation inherent to the EPICA dataset. 

We perform the interpolation using the Akima method, which avoids the spurious oscillations and overshoots that can be introduced by spline-based techniques \cite{akima1970new}, particularly when large gaps are present in the dataset. Previous studies have also shown that interpolation typically does not affect the outcomes of statistical analysis \cite{miller2019testing}. To constrain the Akima interpolation across the 20 kyr gap in the nitrous oxide record between 260 and 240 kyr, we introduce a single artificial data point at 250 kyr by linear interpolation, as proposed by Keyes \textit{et al. }\cite{keyes2023stochastic}. Importantly we note that \citet{salehnia2025continuous} have recently generated a gap-filled time series for N$_2$O by applying machine learning techniques to the EPICA greenhouse gas data. In Appendix \ref{sec:app_C}, we describe how using this gap-filled N$_2$O time series influences our findings.

After interpolation, we removed the slowly varying mean behavior to focus on shorter-timescale variability. To achieve this, we applied a Gaussian smoothing filter with a characteristic time window \textcolor{black}{equal to three times the interpolation interval (approximately 3 kyr). The Gaussian filter provides a smooth estimate of the low-frequency background without introducing the ringing artifacts associated with sharp spectral cutoffs or the discontinuities inherent to moving-average filters. Moreover, the smoothing window chosen is substantially shorter than the dominant Milankovitch periods (23, 41, and 100 kyr), ensuring that we only filter the slowly varying background while preserving the shorter-timescale fluctuations of interest.} 
Subsequently, we subtracted the \textcolor{black}{estimated} mean behavior from the interpolated time series to obtain the fluctuations \textcolor{black}{about} the mean. We then normalized the fluctuation time series, ensuring that each series exhibited a standard deviation of unity, thereby facilitating comparative modeling. The resulting normalized fluctuation time series, shown in Figure \ref{fig:data}b, exhibit approximately Gaussian distributions, supporting our stochastic modeling framework outlined in Sec. \ref{Stochastic_models}.

\section{Multifractal Time-Weighted Detrended Fluctuation Analysis} \label{MFTWDFA}

\subsection{Background}

We employ multifractal time-weighted detrended fluctuation analysis (MFTWDFA) \cite{zhou2010multifractal} to analyze the scaling dynamics and fluctuation structure in the paleoclimate time series. Using this method, we quantify the fluctuations around the mean behavior across various timescales present in the data through a fluctuation function, introduced later in this section. By doing so, we can identify the dominant statistical fluctuations as a function of timescale. When fluctuations in a time series exhibit colored noise, the fluctuation function follows an exponential scaling in time, and the corresponding scaling exponent—referred to as the Hurst exponent—characterizes the color. Consequently, a log-log plot of the fluctuation function will yield a straight line across the range of time where this colored fluctuation behavior occurs, and the slope of this line will represent the corresponding Hurst exponent. Although power spectral analysis can also quantify colored noise, the multifractal approach provides a clearer and more accurate description of the complex multiscale nature of paleoclimate data, particularly in identifying crossover times between distinct noise regimes.

MFTWDFA enhances other existing detrended fluctuation analysis methods, such as MFDFA \cite{kantelhardt2002multifractal}, by incorporating a smoother computation of the mean behavior of the data at each timescale. Whereas MFDFA uses a piecewise polynomial fit to the profile of the data, MFTWDFA employs a time-weighted linear regression within a moving window to provide a continuous estimate of the mean behavior at each timescale. This refinement yields fluctuation functions in which crossover times between distinct noise regimes are more clearly identifiable. In addition, MFTWDFA can resolve fluctuation properties at timescales up to $N/2$ for a dataset of length $N$, in contrast to $N/4$ in MFDFA. Although we have detailed the algorithm for MFTWDFA in our earlier works \cite{agarwal2012trends,agarwal2017exoplanetary,moon2018intrinsic,agarwal2021minimal,keyes2023stochastic}, we outline it here for completeness.

\subsection{Algorithm}

MFTWDFA algorithm involves four steps, as described below:
\begin{enumerate}

\item Given the original time series $X_i$, we first construct a nonstationary profile $Y(i)$ as
\begin{equation}
Y(i) \equiv \sum_{k=1}^i \Big( X_k - \overline{X} \Big), \hspace{0.5cm} i=1,2,...,N,
\label{eqn:profile}
\end{equation}
where $\overline{X}$ is the mean of the time series $X_i$. This profile represents the cumulative deviation of the data from its mean. As mentioned earlier, we require data that is evenly spaced in time. Therefore, we interpolate $Y(i)$ using the Akima method.

\item We divide the nonstationary profile into $N_s = \text{int}(N/s)$ non-overlapping segments of equal length $s$, where $s$ is an integer that varies in the range $1<s \leq N/2$. Each value of $s$ corresponds to a timescale of $s \times \Delta t$, with $\Delta t$ denoting the temporal resolution of the time series under consideration. Because the profile length $N$ is generally not an exact multiple of $s$, we repeat the segmentation procedure starting from the end of the profile and moving backward to the beginning, thereby generating a total of $2N_s$ segments.

\item For each timescale $s$, we detrend the interpolated profile by removing variability on timescales longer than $s$ using a time-weighted linear regression in a window of size $s$. The weights decrease with temporal distance, reflecting the expectation that nearby points are more strongly correlated. As a result, this continuously weighted fit smoothly captures the local mean. We compute the coefficients for the weighted fit, denoted as $\hat{\beta}$, at each point by solving \color{black}
\begin{equation}
\label{eqn:wfit}
     \hat{\beta} = (x^T W x)^{-1} x^T W y,
\end{equation}
where $x$ is the design matrix containing the independent variable (time) and the polynomial basis functions used for the local regression (for a linear fit, a column of ones and the time coordinates within the window), and $y$ is the vector of profile values $Y(i)$ within the same window. \color{black} The elements of the weight matrix $W$ are defined as
\begin{equation}
    w_{ij} = \begin{cases}
    \left(1 - \left(\dfrac{i-j}{s}\right)^2 \right)^2 ,  & |i-j| \leq s \\
    0, & \text{otherwise.}
\end{cases}
\end{equation}
Now, for each timescale $s$, we then compute the variance about the fitted trend, spanning up and down of the the profile, using
\begin{equation}
    V(\nu,s) = \frac{1}{s} \sum_{i=1}^s \Big[ Y([\nu-1]s + i) - \hat{y}([\nu-1]s + i)\Big]^2, \label{eqn:var_up}
\end{equation}
for $\nu = 1,...,N_s,$ and
\begin{equation}
      V(\nu,s) = \frac{1}{s} \sum_{i=1}^s \Big[ Y(N - [\nu - N_s]s + i) - \hat{y}(N - [\nu - N_s]s + i) \Big]^2, \label{eqn:var_down}
\end{equation}
for $\nu = 1,...,N_s,$ where $\nu$ is the index of the moving time window of size $s$ and $\hat{y}(i)$ is the value of the locally weighted linear fit to the profile $Y(i)$.

\item Finally, we compute a generalized fluctuation function, $F_q(s)$, as 
\begin{equation}
\label{eqn:fluct_func}
F_q(s) = \Big[ \dfrac{1}{2N_s}  \sum_{\nu=1}^{2N_s} \{ V(\nu,s)\}  ^{\dfrac{q}{2}} \Big]^{\dfrac{1}{q}},
\end{equation}
where $q$ denotes the statistical moment.
\end{enumerate}

\begin{figure*}
\centering
\includegraphics[width=1.0\textwidth]{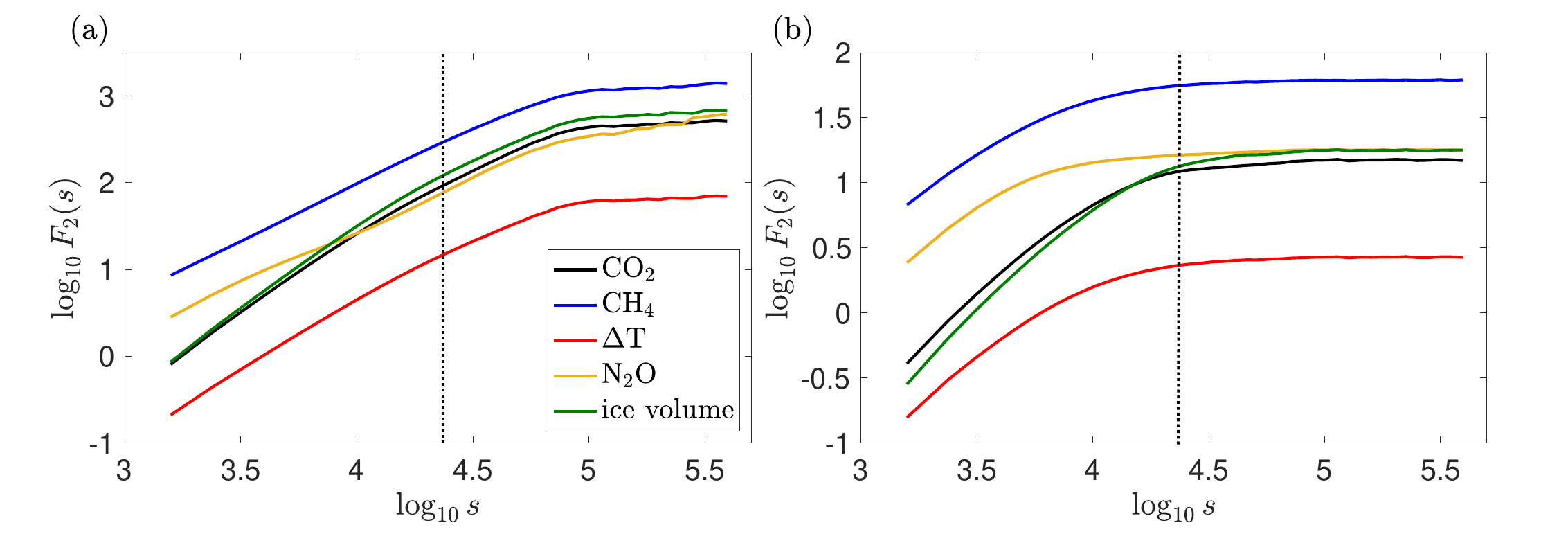}
\caption{Logarithmic plots of the fluctuation functions for (a) original time series and (b) fluctuation time series with the slowly varying behavior removed. The vertical dotted black lines show the 23.5 kyr periodicity used in the later modeling section.}
\label{fig:mftwdfa}
\end{figure*}

\begin{table*}[]
\centering
\large
\renewcommand{\arraystretch}{1.5} 
\begin{minipage}{0.47\textwidth}
\centering
\begin{tabular}{ccc}
\multicolumn{3}{c}{(a)} \\ \hline\hline
& $s < 10^{4.6}$ & $s > 10^{5.1}$ \\ \hline
CO$_2$ & 1.6319 & 0.1375 \\
CH$_4$ & 1.3007 & 0.1600 \\
$\Delta T$ & 1.4820 & 0.1132 \\
N$_2$O & 1.1860 & 0.4907 \\
ice volume & 1.7035 & 0.1562 \\ \hline\hline
\end{tabular}
\end{minipage}
\hspace{0.1cm}
\begin{minipage}{0.47\textwidth}
\centering
\begin{tabular}{ccc}
\multicolumn{3}{c}{(b)} \\ \hline\hline
& $s < 10^{3.8}$ & $s > 10^{5.1}$ \\ \hline
CO$_2$ & 1.6579 & 0.0051 \\
CH$_4$ & 1.1506 & 0.0058 \\
$\Delta T$ & 1.4082 & 0.0055 \\
N$_2$O & 1.2014 & 0.0003 \\
ice volume & 1.7911 & 0.0069 \\ \hline\hline
\end{tabular}
\end{minipage}
\caption{Scaling exponents estimated from MFTWDFA fluctuation function slopes in the two colored-noise regimes for (a) original time series and (b) fluctuation time series. For the shorter-timescale regime, we evaluate slopes for $s < 10^{4.6}$ in the original series and $s < 10^{3.8}$ in the fluctuation series. For the longer-timescale regime, we use $s > 10^{5.1}$ for both sets of time series.}
\label{table:slopes}
\end{table*}

\subsection{Results: Data Analysis}

The dependence of $F_q(s)$ on the timescale $s$ encodes the scaling properties of the time series. For a given statistical moment $q$, the generalized fluctuation function exhibits power-law scaling with $s$, such that $F_q(s) \propto s^{h(q)}$, where $h(q)$ denotes the generalized Hurst exponents. These Hurst exponents provide statistical information about the data at various timescales. If $h(q)$ does not vary with $q$, the time series is characterized as monofractal. In this study, we focus on the second moment ($q=2$), as it directly connects to the power spectrum and provides a simple correspondence between the Hurst exponent $h(2)$ and the noise type \cite{kantelhardt2002multifractal}. For a power spectrum of the form $S(f) \propto f^{-\beta}$, where $f$ is frequency and $\beta$ is the spectral decay exponent, the relationship $h(2) = (1 + \beta)/2$ holds \cite{rangarajan2000integrated}. Consequently, for white noise, characterized by $\beta=0$, it yields $h(2) = 1/2$; in the case of pink noise, where $\beta = 1$, giving $h(2) = 1$; for red noise, with $\beta = 2$, this results in $h(2) = 3/2$, and so forth. This framework allows us to characterize the data in terms of noise type across multiple timescales, where the varying slopes of $\log_{10}F_2(s)$ versus $\log_{10}s$ demonstrate the different dynamical processes operating on various timescales. To demonstrate robust scaling behavior, these linear segments ideally should span as many orders of magnitude as possible; however, the length and sampling resolution of the time series impose upper and lower bounds on the timescales over which to analyze noise behavior.

Figure \ref{fig:mftwdfa}(a) demonstrates that all five variables exhibit similar multiscale stochastic dynamics on timescales both shorter and longer than the 100 kyr glacial-cycle period. Furthermore, the fluctuation functions for the original time series reveal two distinct regimes of colored-noise scaling. To quantify these regimes, we perform linear fits to the log–log fluctuation functions over the intervals $\log_{10}s \in [3.2, 4.6]$ for the original data, $\log_{10}s \in [3.2, 3.8]$ for the fluctuations, and $\log_{10}s \in [5.1, 5.6]$ for the longer-timescale side, and extract the corresponding slopes. Over the timescale range $1.5-40$ kyr (i.e., $10^{3.2}-10^{4.6}$ years), the fitted slopes yield $h(2) \approx 3/2$, corresponding to red-noise behavior. In contrast, over longer timescales of $125-400$ kyr (i.e., $10^{5.1}-10^{5.6}$ years), we observe $h(2)$ values ranging from $0$ to $1/2$, reflecting an anti-persistent colored noise transitioning towards uncorrelated white noise. Notably, the fluctuation function for nitrous oxide deviates slightly from that of the other variables at shorter timescales, revealing subtler crossovers rather than the single slope identified in the other data. Nevertheless, for the modeling purposes, our primary focus remains on the fluctuation function slopes of the data after applying a high-pass filter, which effectively removed the slowly varying behavior.

After applying the high-pass filter described in Sec. \ref{Data_preparation} and then performing MFTWDFA, the resulting fluctuation functions are shown in Figure \ref{fig:mftwdfa}(b). After filtering, the long-timescale slopes decrease toward zero and the crossover shifts to shorter timescales (see Table \ref{table:slopes}). A Hurst exponent of zero implies an absence of long-range scaling and thus indicates mean-reverting behavior of the time series. This behavior arises because the filtered data no longer exhibit a slowly varying background and instead represent a stochastic process that relaxes toward a constant mean. \textcolor{black}{The disappearance of the long-timescale scaling after filtering is therefore expected and confirms that the slowly varying deterministic component has been successfully removed.} As the length of the smoothing window is reduced, higher-frequency components are preferentially removed, causing the observed crossover to shift toward shorter timescales, as shown in Figure \ref{fig:mftwdfa}(b). Importantly, the Hurst exponent for timescales shorter than the smoothing window—approximately 3 kyr—remains unchanged, since the high-pass filter does not suppress variability below the smoothing window. \textcolor{black}{These results underlie our use of a multidimensional non-autonomous Ornstein–Uhlenbeck process to model the data across a range of timescales.  A detailed analysis and further discussion of Fig. \ref{fig:mftwdfa} is given in Appendix \ref{sec:app_A}.}

\section{Stochastic models} \label{Stochastic_models}

\subsection{Background}

Climate time series can be effectively represented using simplified stochastic models, especially when a clear separation exists between short- and long-timescale dynamics, and the short-timescale processes can be treated as random variables \cite{hasselmann1976stochastic}. This approach is of particular interest because it embeds small-scale stochastic fluctuations intrinsic to climate processes within a computationally efficient modeling framework that enables simulations over timescales far longer than those accessible to current global climate models. Moreover, the mathematical simplicity of such models allows for the analytical estimation of key parameters, such as system stability and noise amplitude, and facilitates the incorporation of coupling functions that provide insight into the mechanisms governing interactions between variables within the climate system.

The results obtained from MFTWDFA support the applicability of a stochastic modeling framework for analyzing the climate variables and ice volume time series during the Late Pleistocene. For all five time series, the contrast between short- and long-timescale dynamics reveals a pronounced separation of timescales between sub-glacial and super-glacial periods. Moreover, following the removal of long-term glacial-cycle variability, the remaining short-term sub-glacial signal is nonstationary and exhibits approximately red noise characteristics.

The stochastic model employed here builds upon the Ornstein–Uhlenbeck process, extending its applicability to a non-autonomous system characterized by a distinct separation of timescales. The Ornstein–Uhlenbeck process is the overdamped limit of the Langevin equation for a Brownian particle in a quadratic potential, yielding a linear mean-reverting stochastic differential equation driven by Gaussian white noise. To account for the slowly varying mean behavior of the glacial–interglacial cycles, we can add a longer-timescale forcing term to this model, which modulates the system's mean state. Therefore, the resulting model can appropriately represent both the short-term fluctuations and the long-term, slowly evolving mean behavior of the paleoclimate time series.

We aim to model the prominent Milankovitch frequencies present in the paleoclimate time series; therefore, our model coefficients are time-dependent and periodic. This periodicity allows us to determine the model coefficients from the periodic statistics of the data. The period of the coefficients based on (a) the power spectra of the time series, (b) the periodicities associated with Milankovitch cycles, and (c) the timescale-separated noise structure revealed by MFTWDFA. To identify a common spectral peak across the five datasets—which also corresponds to a Milankovitch period—we compare their power spectra (Fig. 2) in order to locate this peak. We then examine the MFTWDFA fluctuation functions (Fig. 3) to verify that this peak occurs at a sufficiently short timescale for the dynamics to exhibit approximately red-noise behavior. We find that the most dominant such frequency is approximately $4.25 \times 10^{-5}$ year$^{-1}$, corresponding to a period of about 23.5 kyr, which corresponds to the Milankovitch cycle associated with the combined effects of axial and apsidal precession \cite{hays1976variations}.

\subsection{One-variable model} \label{Stochastic_models_1D}

Following our previous study \cite{keyes2023stochastic}, we treat the dynamics of an individual time series with a one-dimensional non-autonomous Ornstein–Uhlenbeck model. The evolution of the $i$th variable, denoted as $\eta_i(t)$, is described by;
\begin{equation}
\frac{d\eta_i(t)}{dt} = a_i(t) \eta_i(t) + N_i(t)\xi_i(t). \label{eqn:1D_model}
\end{equation}
The time-periodic deterministic term, $a_i(t)$, drives the mean-reverting drift and characterizes the stability of the system. Specifically, negative (positive) values of $a_i(t)$ correspond to stable (unstable) dynamics, leading to exponential decay (growth) of fluctuations. The noise amplitude, $N_i(t)$, is also defined as a deterministic periodic function, representing temporal modulation of the stochastic forcing, $\xi_i(t)$, which denotes Gaussian white noise characterized by the properties $\langle \xi_i(t) \rangle=0$ and $\langle \xi_i(t) \xi_i(t') \rangle = \delta (t-t')$, where $\langle \cdot \rangle$ is the time average and $\delta$ is the Dirac delta function.

\textcolor{black}{The white-noise forcing represents the combined influence of numerous unresolved climate processes, including internal atmospheric and oceanic variability, biogeochemical fluctuations, and measurement uncertainty. Since these processes act on timescales much shorter than those recorded in paleoclimate data, their cumulative impact can be effectively approximated as Gaussian white noise by the central limit theorem. The MFTWDFA analysis presented in Sec. \ref{MFTWDFA} and analyzed in Appendix \ref{sec:app_A} demonstrates the importance of treating the system with a non-autonomous Ornstein–Uhlenbeck process, which quantifies the general influence of fluctuations about a background state captured by the periodic coefficients accounting for the slow modulation induced by orbital forcing.  The model exhibits distinct dynamical behavior depending on whether timescales are much shorter than, comparable to, and much longer than the modulation period $P=23.5~\mathrm{kyr}.$   For example, after removing the slowly varying mean behavior, the residual fluctuations exhibit mean-reverting behavior at long timescales, while their probability distributions are approximately Gaussian.  Moreover, on short timescales, the statistical characteristics are consistent with an autonomous Ornstein–Uhlenbeck process, in which a linear restoring mechanism transforms Gaussian white-noise forcing into a red-noise process.  The framework provides a coherent means of analyzing the observed multi-scale temporal variability in the filtered paleoclimate records.}


We determine the time-dependent coefficients $a_i(t)$ and $N_i(t)$ using a modified version of the methodology introduced by \citet{moon2017unified}. Further details regarding this approach can be found in the supplementary information provided in their work. \textcolor{black}{Our analysis considers a time series consisting of $M=34$ periods, each of length $P \approx 23.5$ kyr (corresponding to the dominant net precession cycle), discretized into $T=25$ points per period.} In this framework, the periodic drift coefficient in Eq. (\ref{eqn:1D_model}) is defined as $a_i(t)=a_i([t/\Delta t] \,\textrm{mod} \,T)$, where $[ \cdot ]$ is the integer part and $\Delta t = P/T \approx 0.94$ kyr. We also employ a similar periodic representation for the noise amplitude coefficient $N_i(t)$. We determine the discrete set of coefficients $a_i(k)~\forall~ k \in [1,T]$ using the analytic solution of the stochastic model, yielding
\begin{equation}
\label{eqn:a_coeff}
a_i(k) \approx - \frac{1}{\Delta t} \frac{S_i(k) - A_i(k)}{S_i(k)},
\end{equation}
where $S_i(k)$ and $A_i(k)$ denote the approximate periodic variance and autocorrelation of the observed time series $X_i$, respectively, and are defined as
\begin{align}
S_i(k) &\equiv  \frac{1}{M-1} \sum_{j=0}^{M-1} X^{j T+k}_i X^{j T+k}_i \approx \langle (\eta_i(k))^2 \rangle, \qquad \textrm{and} \\
A_i(k) &\equiv \frac{1}{M-1} \sum_{j=0}^{M-1} X^{j T+k}_i X^{j T+k+1}_i \approx \langle \eta_i(k)\eta_i(k+1) \rangle.
\end{align}

Finally, by combining the above formula for $a_i(k)$ with Eq. (\ref{eqn:1D_model}), the expression for $N_i(k)$ follows as
\begin{equation}
{N_i(k)} = \sqrt{\frac{\langle y_i^2(k) \rangle}{\Delta t}},
\end{equation}
where $y_i(k)$ is the stochastic residual of the discretized form of Eq. (\ref{eqn:1D_model}), and is given by 
\begin{equation}
y_i(k) \equiv \eta_i(k+1) - \eta_i(k) - a_i(k)\eta_i(k)\Delta t. 
\end{equation}

\subsection{Five-variable model} \label{Stochastic_models_5D}

We extend the one-dimensional framework to a multivariate setting in order to analyze multiple time series simultaneously. \textcolor{black}{Climate variables can exhibit substantial shared variability arising from common external forcing and mutual interactions within the climate system. Importantly, the multivariate formulation can extract dynamical coherence among the datasets by requiring their evolution to be described consistently within a common system of coupled stochastic equations. By treating all variables within a single coupled framework, the model attributes the evolution of each variable not only to its intrinsic dynamics and stochastic forcing, but also to the coupled influence of the other variables. Consequently, variability shared across the records can be represented by their coupling structure rather than interpreted independently in separate univariate models.} By introducing linear coupling, we represent the mutual influences among the variables \textcolor{black}{and obtain} first-order estimates of the dominant directions \textcolor{black}{and strengths} of \textcolor{black}{these interactions}. \textcolor{black}{This provides dynamical information beyond that contained in} symmetric \textcolor{black}{statistical} measures, such as covariance, \textcolor{black}{which quantify shared variability but do not distinguish the direction of influence.} Following the approach of \citet{moon2019coupling} and Keyes \textit{et al.} \cite{keyes2023stochastic}, we construct a five-variable model for the $\text{CO}_2$, $\text{CH}_4$, $\text{N}_2\text{O}$, temperature, and ice volume time series, along with their pairwise couplings. We emphasize that this modeling framework is readily extendable to systems with an arbitrary number of variables.

The system of five coupled stochastic differential equations can be written as
\begin{equation}
    \frac{d\eta_i(t)}{dt} = a_i(t)\eta_i(t) + N_i(t)\xi_i(t) + \sum_{j \neq i} b_{ij}(t) \Big[ \eta_j(t) - \eta_i(t) \Big], \label{eqn:5D_model}
\end{equation}
where $b_{ij}(t)$ is a linearized diffusive coupling term that captures the influence of $\eta_{j}(t)$ on $\eta_i(t)$. This form of coupling, expressed as the difference between interacting variables, has been employed in a broad range of systems (see for example \citet{othmer1971instability}, \citet{levin1974dispersion}, \citet{kopell1973plane}, and \citet{krause2021modern}, among many others).

In the multivariate formulation, the coefficients $a_i(k)$ and $b_{ij}(k)$ can be obtained directly by solving a set of linear matrix equations. Specifically, we construct five independent systems—one for each model equation—by multiplying the equation for $\eta_i(t)$ by each of the variables $\eta_i(t)$ and then taking the ensemble average. This procedure yields the following linear system of equations
\begin{eqnarray}
&\begin{bmatrix}
E_{ii} & E_{ij} - E_{ii} & E_{ik} - E_{ii} & E_{il} - E_{ii} & E_{im} - E_{ii} \\
E_{ij} & E_{jj} - E_{ij} & E_{jk} - E_{ij} & E_{jl} - E_{ij} & E_{jm} - E_{ij} \\
E_{ik} & E_{jk} - E_{ik} & E_{kk} - E_{ik} & E_{kl} - E_{ik} & E_{km} - E_{ik} \\
E_{il} & E_{jl} - E_{il} & E_{kl} - E_{il} & E_{ll} - E_{il} & E_{lm} - E_{il} \\
E_{im} & E_{jm} - E_{im} & E_{km} - E_{im} & E_{lm} - E_{im} & E_{mm} - E_{im} 
\end{bmatrix}& \nonumber \\
&\times
\begin{bmatrix}
a_i(t) \\
b_{ij}(t) \\
b_{ik}(t) \\
b_{il}(t) \\
b_{im}(t) 
\end{bmatrix}
=
\begin{bmatrix}
D_{ii} \\
D_{ij} \\
D_{ik} \\
D_{il} \\
D_{im} 
\end{bmatrix}, &
\end{eqnarray}
where $E_{xy} = \langle \eta_x(t) \eta_y(t) \rangle$ denotes the covariance between variables $x$ and $y$, and $D_{xy} = \langle \frac{d\eta_x}{dt} \eta_y(t) \rangle$ represents the corresponding cross-moment involving the time derivative. By solving these matrix systems, we can determine the drift and coupling coefficients for each variable.

To estimate the noise-amplitude coefficient $N_i(k)$, we multiply each equation by its corresponding $\eta_i(t+\Delta t)$ and then take the ensemble average, which leads to
\begin{align}
        N_{i}^2(t) = & \Big\langle \eta_{i}(t + \Delta t) \frac{d\eta_{i}(t)}{dt} \Big\rangle - a_{i}(t)\langle \eta_{i}(t) \eta_{i}(t + \Delta t)\rangle \nonumber \\ & - \sum_{j \neq i} b_{ij}(t) \Big[ \langle \eta_{i}(t + \Delta t) \eta_{j}(t) \rangle - \langle \eta_{i}(t + \Delta t) \eta_{i}(t) \rangle \Big].
\end{align}

\subsection{Results: Model} \label{Result: model}

We apply this modeling approach to Late Pleistocene paleoclimate datasets to derive and interpret the stability, coupling, and noise coefficients for each of the five variables in the coupled system. In this subsection, we present the model coefficients computed using the EPICA record. In Appendix \ref{sec:app_C}, we use the gap-filled N$_2$O data referred to in \S \ref{Data_preparation} and compute the resulting model coefficients.

\begin{figure*}
\centering
\begin{subfigure}[b]{1.0\textwidth}
    \includegraphics[width=\linewidth]{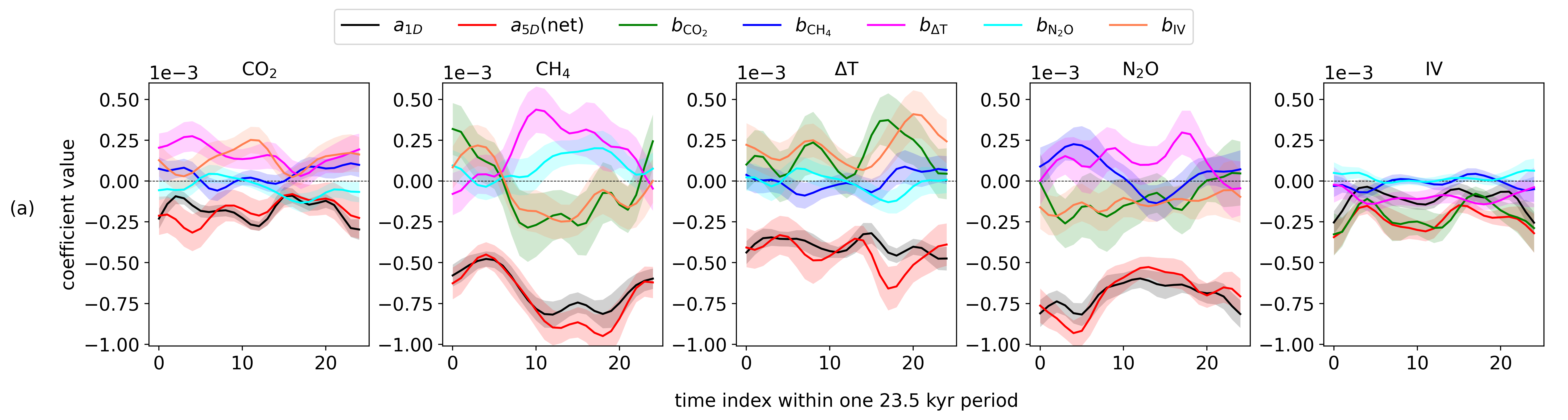}
\end{subfigure}
\hfill
\begin{subfigure}[b]{1.0\textwidth}
    \includegraphics[width=\linewidth]{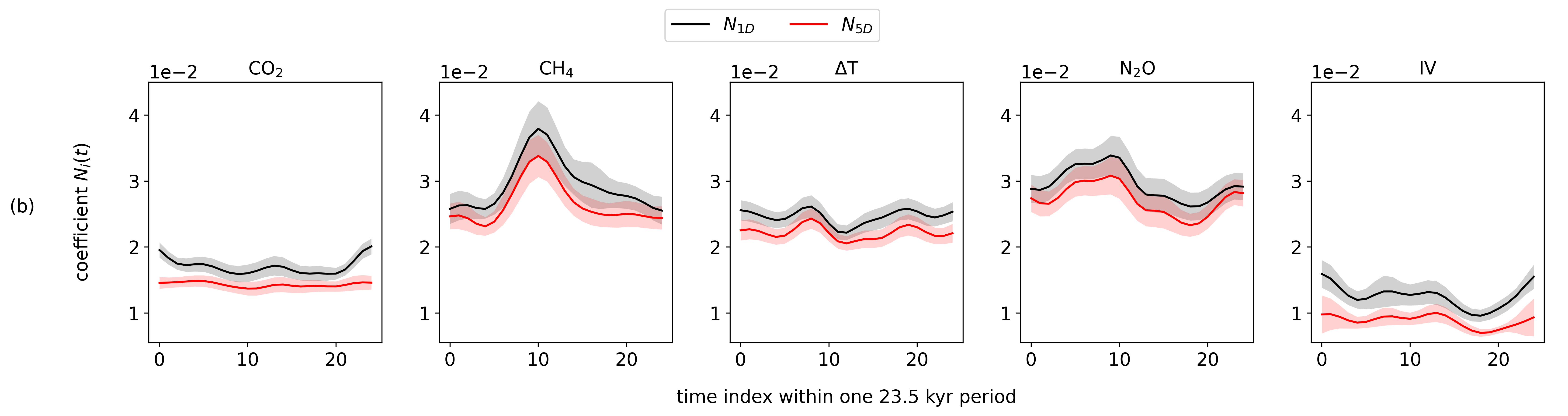}
\end{subfigure}
\caption{\textcolor{black}{Time-dependent model coefficients over one $P \approx 23.5$ kyr cycle. The horizontal axis denotes the elapsed time within this coefficient cycle, which is discretized into $T=25$ equally spaced points ($\Delta t \approx 0.94$ kyr).} (a) Stability coefficient for the one-dimensional model and the net stability coefficient for the five-dimensional model, defined as $a_{i,5D}$(net) $= a_{i,5D} - \sum_{j} b_{ij}$, and coupling coefficients for the five-dimensional model. In the legend, the generic symbol $b_{x}$ represents the influence of variable $x$ on the other variables. (b) Noise amplitude coefficients for one- and five-dimensional models. \textcolor{black}{In both panels, the solid curves denote the coefficient estimates obtained from the complete dataset, while the shaded regions represent the corresponding bootstrap standard error 95\% confidence intervals.}}
\label{fig:model_coefficients}
\end{figure*}

\textcolor{black}{To quantify the statistical robustness of the inferred time-dependent coefficients, we performed a period-wise bootstrap analysis. The original dataset consists of $M=34$ complete Milankovitch-scale periods. Each bootstrap realization is generated by resampling these 34 periods with replacement while preserving the 25 phase points within every selected period. The same resampled period indices are used simultaneously for all five variables, thereby preserving the covariance structure among $\Delta \mathrm{T}, \mathrm{CO}_2$, $\mathrm{CH}_4$, $\mathrm{N}_2 \mathrm{O}$ and IV (ice volume). For every bootstrap realization, we computed $a_{1 D}(t), N_{1 D}(t)$, $a_{5 D}(t), N_{5 D}(t)$, and $b_{i j}(t)$. The bootstrap standard error is computed from the distribution of these estimates. In Figure \ref{fig:model_coefficients}, solid curves denote the estimates from the full dataset, while the shaded regions represent the standard error with 95\% bootstrap confidence intervals, providing a direct visualization of finite-sample uncertainty associated with having only 34 independent periods.
}

\subsubsection{Deterministic stability coefficients} \label{Deterministic_stability_coefficients}

In the one-variable system [Eq. \eqref{eqn:1D_model}], the deterministic stability is governed solely by the drift coefficient $a(t)$, whereas in the five-variable system [Eq. \eqref{eqn:5D_model}], the corresponding net deterministic stability is given by $a_{i,5D}$(net) $= a_{i,5D} - \sum_{j} b_{ij}$. Figure \ref{fig:model_coefficients}(a) shows the deterministic stability coefficients for the one- and five-dimensional models. It is important to emphasize that these coefficients are comparable across models for each variable, as they pertain to the same fundamental processes. However, the five-variable model shows slightly more negative values for all five variables, suggesting that inter-variable couplings enhance overall stability. In both models, methane and nitrous oxide exhibit greater stability compared to carbon dioxide, temperature and ice volume (IV), implying that their deterministic drift exerts a stronger influence, driving them toward the long-term mean behavior.

\subsubsection{Coupling coefficients} \label{Coupling_coefficients}

Figure \ref{fig:model_coefficients}(a) shows the time-dependent coupling coefficients for the five-variable model. The sign and magnitude of these coefficients characterize the dynamical interactions among the variables involved. By construction, the coupling coefficient $b_{ij}(t)$ ($b_{ji}(t)$), which connects variable $j$ to variable $i$, or symbolically $j \rightarrow i$, represents the influence exerted by variable $j$ on variable $i$ (variable $i$ on variable $j$). Thus, $|b_{ij}(t)|$ measures the interaction strength, and the sign of $b_{ij}(t)$ determines the direction of the influence. For example, a positive (negative) value of $b_{ij}(t)$ indicates that variable $j$ suppresses (enhances) the growth of variable $i$. In stability terminology—i.e., local time decay versus growth of perturbations—positive (negative) coupling coefficients have a stabilizing (destabilizing) effect on the coupled variables. As an illustrative case, when $b_{ij}(t) > 0$ and $b_{ji}(t)<0$, it follows that variable $j$ suppresses the growth of variable $i$, while simultaneously, variable $i$ promotes the growth of variable $j$.

The sign of the coupling coefficients shown in Figure \ref{fig:model_coefficients}(a) plays a role in the intervariable stability as discussed above, but the overall interactions also depend on both their time-dependence and magnitude relative to other coefficients in the theory.  In this paragraph, we mainly discuss the sign.
The coupling coefficients associated with $\Delta \text{T} \rightarrow \text{IV}$ and $\text{CO}_2 \rightarrow \text{IV}$ are negative, so that temperature and carbon dioxide facilitate the growth of ice volume, whereas those associated with $\text{IV} \rightarrow \Delta \text{T}$ and $\text{IV} \rightarrow \text{CO}_2$ are positive, indicating that ice volume exerts a suppressive-controlling effect on these two variables.  This reflects the canonical observation of the co-variation of these three quantities through glacial and interglacial periods (Fig. \ref{fig:data}(a)), with their associated phase shifts determined by the time dependence of the coefficients. While the coupling coefficients for $\text{CO}_2 \rightarrow \text{CH}_4$ and $\text{CO}_2 \rightarrow \text{N}_2\text{O}$
are predominantly negative, the reverse relationships have coefficients of both signs with relatively small magnitudes. The coefficients for $\Delta \text{T} \rightarrow \text{CH}_4$ and $ \Delta \text{T} \rightarrow \text{N}_2\text{O}$ are primarily positive, whereas the reverse relationships have both positive and negative coefficients with small magnitudes. The coupling coefficients for $\text{CH}_4 \rightarrow \text{IV}$ and $\text{N}_2\text{O} \rightarrow \text{IV}$ oscillate around the origin, implying that methane and nitrous oxide have a weak influence on ice volume, whereas the reverse relationships $\text{IV} \rightarrow \text{CH}_4$ and $\text{IV} \rightarrow \text{N}_2\text{O}$ have, on average, negative coupling coefficients, so that an increase in ice volume tends to increase the concentrations of these greenhouse gases in the absence of other effects which, as discussed below in \S \ref{Response_functions}, dominate and have the opposite effect.  In Appendix \ref{sec:app_B}, we provide a simple approximate example of the influence of weak coupling of methane and nitrous oxide to ice volume. The generally larger magnitudes of the coupling coefficients associated with carbon dioxide, temperature, and ice volume underscore their important influence on the overall dynamics, although the detailed trajectories ultimately arise from the interactions among all five variables. 

\textcolor{black}{The comparison of the present five-variable model with the four-variable model of  Keyes \textit{et al.} \cite{keyes2023stochastic} indicates that the signs and temporal structure of the dominant couplings among $\Delta$T and CO$_2$ remain broadly consistent, although their magnitudes exhibit quantitative changes when IV is explicitly incorporated. This behavior is expected in a reduced stochastic description, since interactions mediated by the cryosphere that were previously absorbed into the effective dynamics of the four-variable system are now represented explicitly through the IV coupling terms. Thus, the persistence of the principal coupling structure demonstrates the robustness of the earlier results, whereas the quantitative changes and the newly resolved interactions with IV characterize the additional dynamics introduced here.}

\subsubsection{Noise amplitude coefficients} \label{Noise_amplitude_coefficients}

The periodic behavior of the noise terms in both the one-variable and five-variable models shows similar dynamics across all variables [see Figure \ref{fig:model_coefficients}(b)]. However, the noise amplitude associated with the five-variable model is consistently lower than that of the corresponding one-variable model. \textcolor{black}{This difference can be interpreted within the framework of stochastic model reduction \cite{hasselmann1976stochastic}. In the one-variable model, the effects of the omitted climate variables and their interactions remain unresolved and are therefore effectively incorporated into the stochastic forcing. In contrast, the five-variable model explicitly represents part of this variability through the inter-variable coupling terms.} Therefore, each variable in the five-variable model compensates by contributing a smaller amount of noise, thereby sustaining a comparable overall noise level across the models.

\subsection{Model interpretation} \label{Model_interpretation}

\begin{figure*}
\centering
\begin{subfigure}[b]{1.0\textwidth}
    \includegraphics[width=\linewidth]{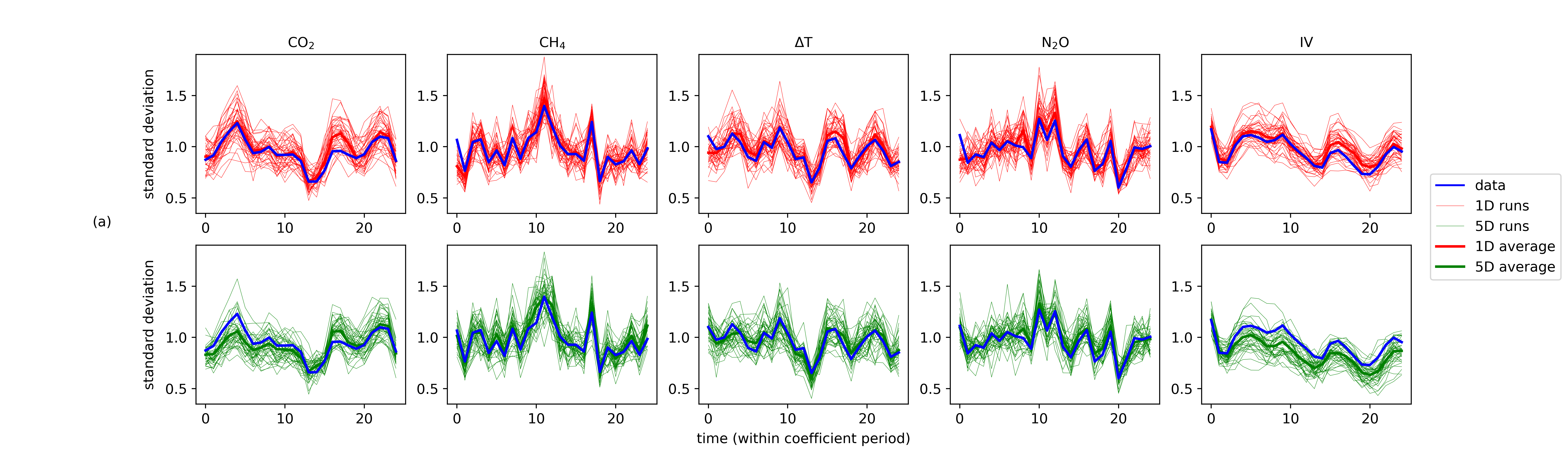}
\end{subfigure}
\hfill
\begin{subfigure}[b]{1.0\textwidth}
    \includegraphics[width=\linewidth]{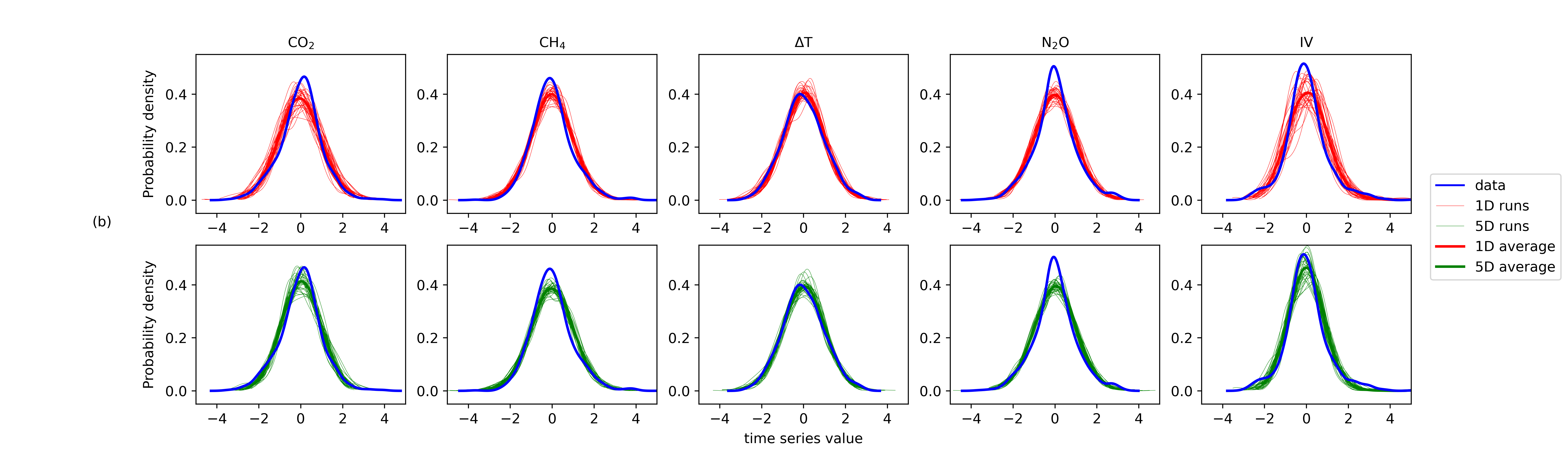}
\end{subfigure}
\hfill
\begin{subfigure}[b]{1.0\textwidth}
    \includegraphics[width=\linewidth]{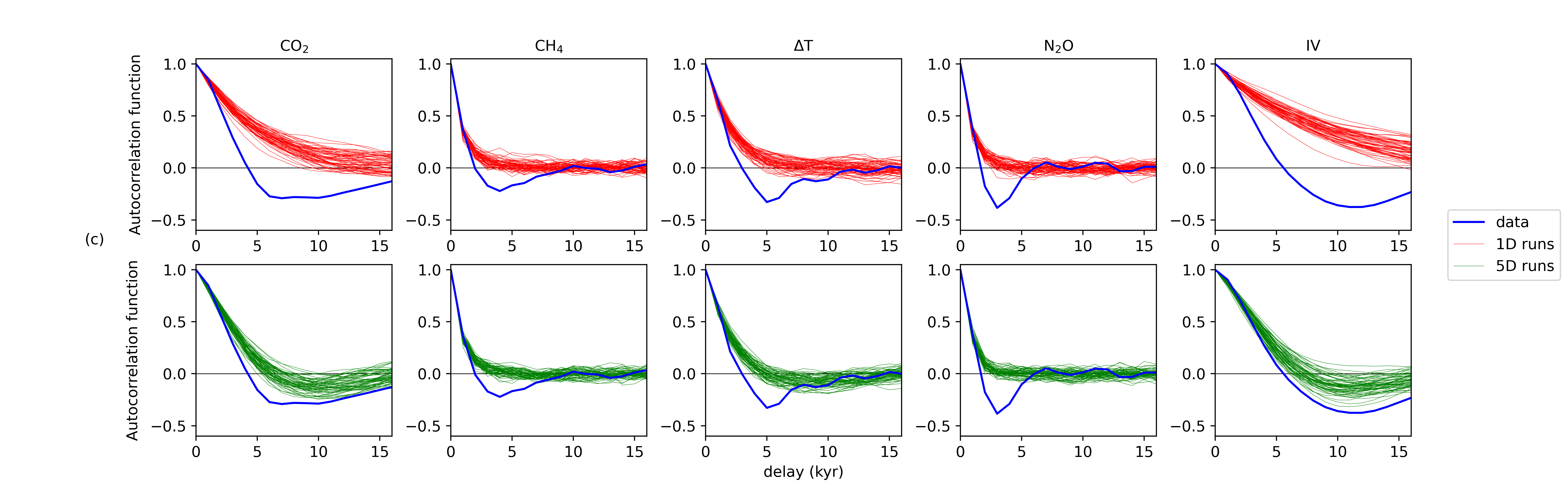}
\end{subfigure}
\caption{Comparison of (a) periodic standard deviations, (b) probability density functions, and (c) autocorrelation functions between the data and one-variable (top rows) and five-variable (bottom rows) models.}
\label{fig:model_fidelity}
\end{figure*}

The coupling structure among temperature, ice volume, and greenhouse gases reveals their mutual interactions and elucidates the dynamics governing Late Pleistocene climate variability. The negative coupling coefficients connecting $\Delta$T and CO$_2$ to ice volume imply that these variables have a destabilizing influence on ice volume. Surface air temperature plays a key role in regulating the mass balance of ice sheets by affecting accumulation and ablation; cooling promotes ice growth and warming accelerates ice retreat \cite{roe2006defense,shakun2015800}.  Similarly, variations in atmospheric CO$_2$ exert a strong greenhouse gas forcing, thereby influencing ice-sheet mass balance: lower CO$_2$ concentrations promote ice growth, while higher concentrations enhance ice retreat. Paleoclimate records consistently show low atmospheric CO$_2$ during glacial periods and higher levels during interglacial periods \cite{bereiter2015revision,berends2021reconstructing}. \textcolor{black}{These newly resolved $\Delta \mathrm{T} \rightarrow \mathrm{IV}$ and CO$_2 \rightarrow \mathrm{IV}$ pathways are not available in the four-variable model, and provide a direct dynamical route through which thermal and radiative forcing can affect the slower cryospheric component of the climate system.} 

Ice sheets influence temperature through many mechanisms, including the ice–albedo feedback, elevation-induced cooling, and large-scale modifications to atmospheric and oceanic circulation \cite[e.g.,][]{SaltzmanMaasch1988, Alley1999,bintanja2005modelled,abe2013insolation,booth2024ice}. These processes amplify cooling during glacial periods and reinforce warming during deglaciation. Changes in ice volume also affect atmospheric CO$_2$ by altering ocean circulation and ventilation, influencing CO$_2$ solubility due to variations in sea surface temperature, and impacting the stratification and storage of CO$_2$ in the ocean as sea levels change \cite{ganopolski2009nature,kurahashi2010effects,shakun2012global,shakun2015800}. These dynamics enhance deep-ocean carbon sequestration during glacial periods and facilitate rapid CO$_2$ release during deglaciation.  These processes are reflected in the positive coupling coefficients linking ice volume to $\Delta$T and CO$_2$ shown in Figure \ref{fig:model_coefficients}(b). \textcolor{black}{In the earlier four-variable formulation, these feedbacks could only indirectly influence $\Delta \mathrm{T}$ and CO$_2$, and were therefore incorporated into effective atmospheric coupling coefficients and residual noise. Their explicit representation in the present model redistributes part of that effective interaction into the $\mathrm{IV} \rightarrow\Delta \mathrm{T}$ and $\mathrm{IV} \rightarrow \mathrm{CO}_2$ terms. This distinction is especially important because the opposite signs of the forward and reverse IV couplings introduce delayed cryosphere-climate feedbacks and permit damped oscillatory response modes that were not reproduced in the earlier model (see Sec. \ref{Response_functions} for details).} 

The model captures the asymmetric interactions between ice volume and the greenhouse gases CH$_4$ and N$_2$O. Both gases inhibit ice-sheet growth through their greenhouse warming effects \cite{loulergue2008orbital,schilt2010glacial}. In turn, ice sheets affect the atmospheric concentrations of CH$_4$ and N$_2$O by modifying boundary conditions that regulate their primary source processes, such as wetland extent, hydrology, ocean circulation, and ocean ventilation \cite{schmittner2008glacial,hopcroft2017understanding}. Additionally, the strong cooling induced by ice sheets suppresses microbial processes involved in methane and nitrogen cycling, whereas deglacial warming and associated changes in ocean circulation promote rapid increases in CH$_4$ and N$_2$O concentrations. \textcolor{black}{The present five-variable formulation, therefore, separates direct greenhouse gas interactions from variability mediated through the cryosphere, which is possible because IV is included as an explicit state variable.} 

Finally, our model captures the mutually stabilizing coupling between temperature and carbon dioxide, congruent with phase analyses that show a close, yet state-dependent, $\Delta\text{T}-\text{CO}_2$ coupling across glacial cycles \cite{shackleton2000100,shakun2012global,shakun2015800}. The model further captures a predominantly positive feedback of $\Delta$T on $\text{CH}_4$ and $\text{N}_2\text{O}$. This is consistent with evidence that warming enhances methane–climate feedbacks through an increase in emissions from wetland and permafrost and changes in atmospheric methane lifetime \cite{cheng2022impact}, as well as an increase in $\text{CH}_4$ and $\text{N}_2\text{O}$ emissions from the terrestrial biosphere under warming conditions \cite{arneth2010terrestrial,van2011increased}. In contrast, the model captures a predominantly negative feedback of CO$_2$ on $\text{CH}_4$ and $\text{N}_2\text{O}$, which contradicts the findings of Keyes \textit{et al.} \cite{keyes2023stochastic}. This discrepancy may stem from the dominant effects of ice sheets, which were not considered in their study. Both $\text{CH}_4$ and $\text{N}_2\text{O}$ can exert a strong impact on climate due to their high radiative efficiencies, such that even minor perturbations in their emissions can result in amplified temperature responses \cite{khalil1989climate}.

\subsection{Model fidelity} \label{Model_fidelity}

We use the model coefficients computed from the paleoclimate records to integrate the one- and five-variable models [Eq. \eqref{eqn:1D_model} and Eq. \eqref{eqn:5D_model}] forward in time using a standard Euler scheme. This process generates synthetic time series whose statistical properties and noise characteristics should match those of the original detrended data. To assess how effectively the simulations capture the various features of the observed data, we compare key statistical metrics.

Figure \ref{fig:model_fidelity}(a) presents a comparison of the periodic standard deviations of the data and those produced by the models for each variable. Both the one- and five-variable models reproduce the overall amplitude and periodic structure of the standard deviation across all time series. The probability distribution functions also show good overall agreement; however, as illustrated in Figure \ref{fig:model_fidelity}(b), the observed distributions exhibit slight departures from Gaussianity, \textcolor{black}{particularly for small fluctuation amplitudes near the central part of the distributions, whereas the Ornstein–Uhlenbeck model with Gaussian forcing produces approximately Gaussian statistics. These minor deviations indicate the presence of statistical structure in the observations that is not fully captured by the Gaussian approximation and may reflect untreated nonlinear dynamics  or the influence of additional processes or variables in the observed system that are not accounted for in our models.} Among the five variables examined, ice volume shows the most non-Gaussian characteristics and is therefore the least accurately captured by our modeling framework.

Finally, in Figure \ref{fig:model_fidelity}(c), we present a comparison of the autocorrelation functions derived from the data and the model simulations, which reveals larger discrepancies than those observed for the other statistical measures. \textcolor{black}{This is not unexpected, as the autocorrelation function provides a more stringent test of the model than the variance or probability distribution because it probes the temporal memory and lag-dependent structure of the fluctuations rather than their one-point statistics. Therefore, reproducing the autocorrelation requires the model to treat not only the amplitude and distribution of the fluctuations, but also their temporal persistence and relaxation dynamics.} Although both the one- and five-variable models capture some of the oscillatory structure present in the autocorrelation functions of all five variables, neither model accurately reproduces the depth of the negative minimum observed in the data or the rate at which the autocorrelation decays toward that minimum. Here again, in addition to inherent model simplifications that may not fully account for the nonlinear dynamics of the paleoclimate system, these discrepancies may arise from the influence of additional processes or variables that interact with the modeled variables in the real climate system but are not explicitly included in the model framework. \textcolor{black}{They may also reflect temporal dependencies beyond, for example, those represented by our particular non-autonomous framework, or colored or correlated stochastic forcing. Furthermore, the empirical autocorrelation functions are estimated from a finite number of periods ($M=34$), which introduces sampling uncertainty that becomes increasingly relevant at larger time lags.} Notably, the five-variable model reproduces the decay rate and some negative autocorrelation values better than the one-variable model, highlighting the role of inter-variable coupling in shaping temporal dependencies. Nevertheless, we regard the limited ability to reproduce the autocorrelation structure as a key limitation of the predictive capability of our modeling approach.

\textcolor{black}{While Figure \ref{fig:model_fidelity}(c) focuses on the autocorrelation of each variable, the temporal correlations between different variables provide complementary information about their joint dynamics. The response-function analysis presented in the following subsection builds directly on these cross-correlations, using the full time-dependent correlation matrix to quantify the directional relationships between the variables.}

\subsection{Response functions} \label{Response_functions}

\begin{figure*}
\centering
\includegraphics[width=1.0\textwidth]{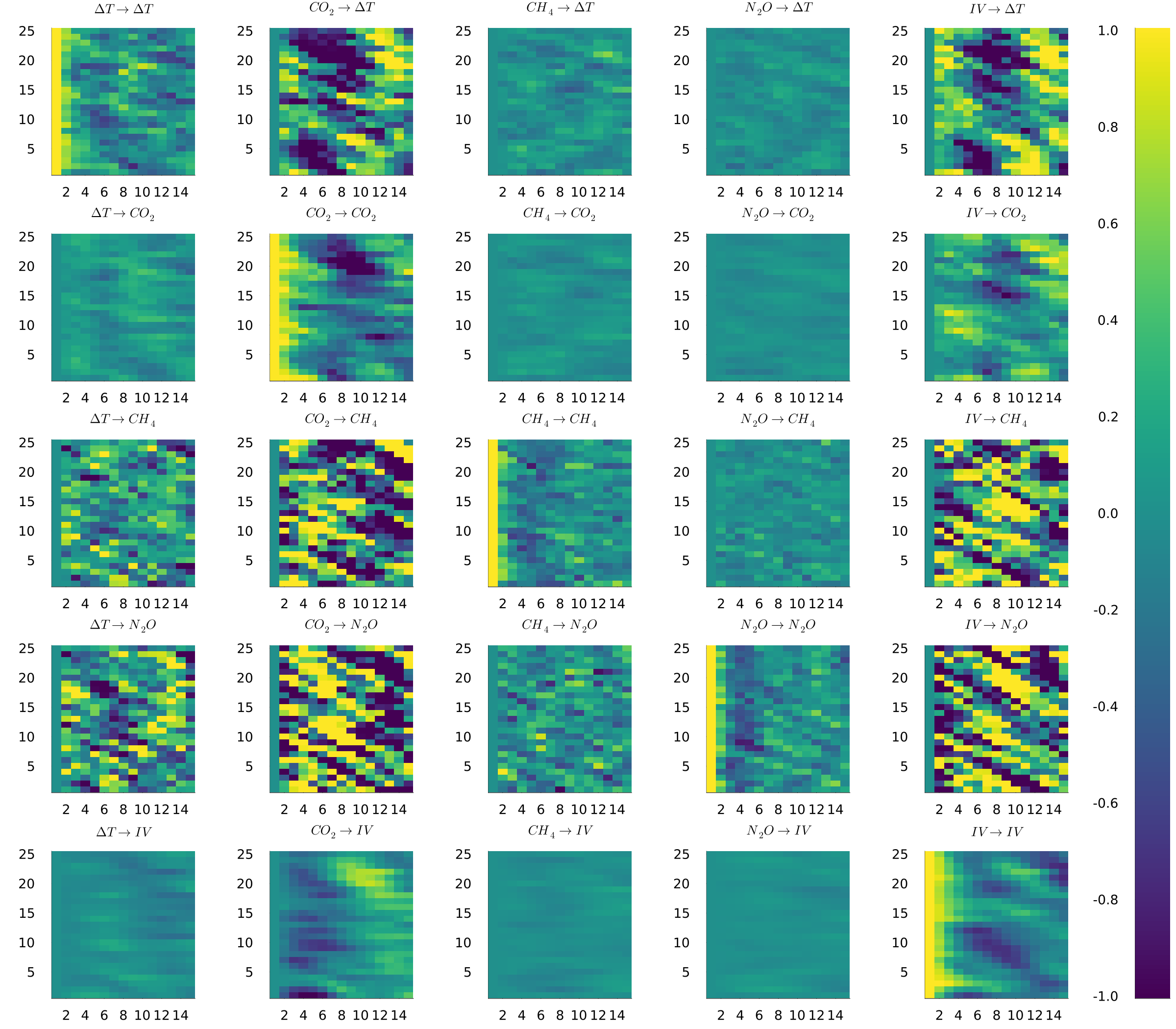} 
\caption{Matrix elements of the time periodic response function obtained from the data plotted as a function of time. The response function is given by Eq. \eqref{Response_function_data}.}
\label{fig:Response_function_data}
\end{figure*}

\begin{figure*}
\centering
\includegraphics[width=1.0\textwidth]{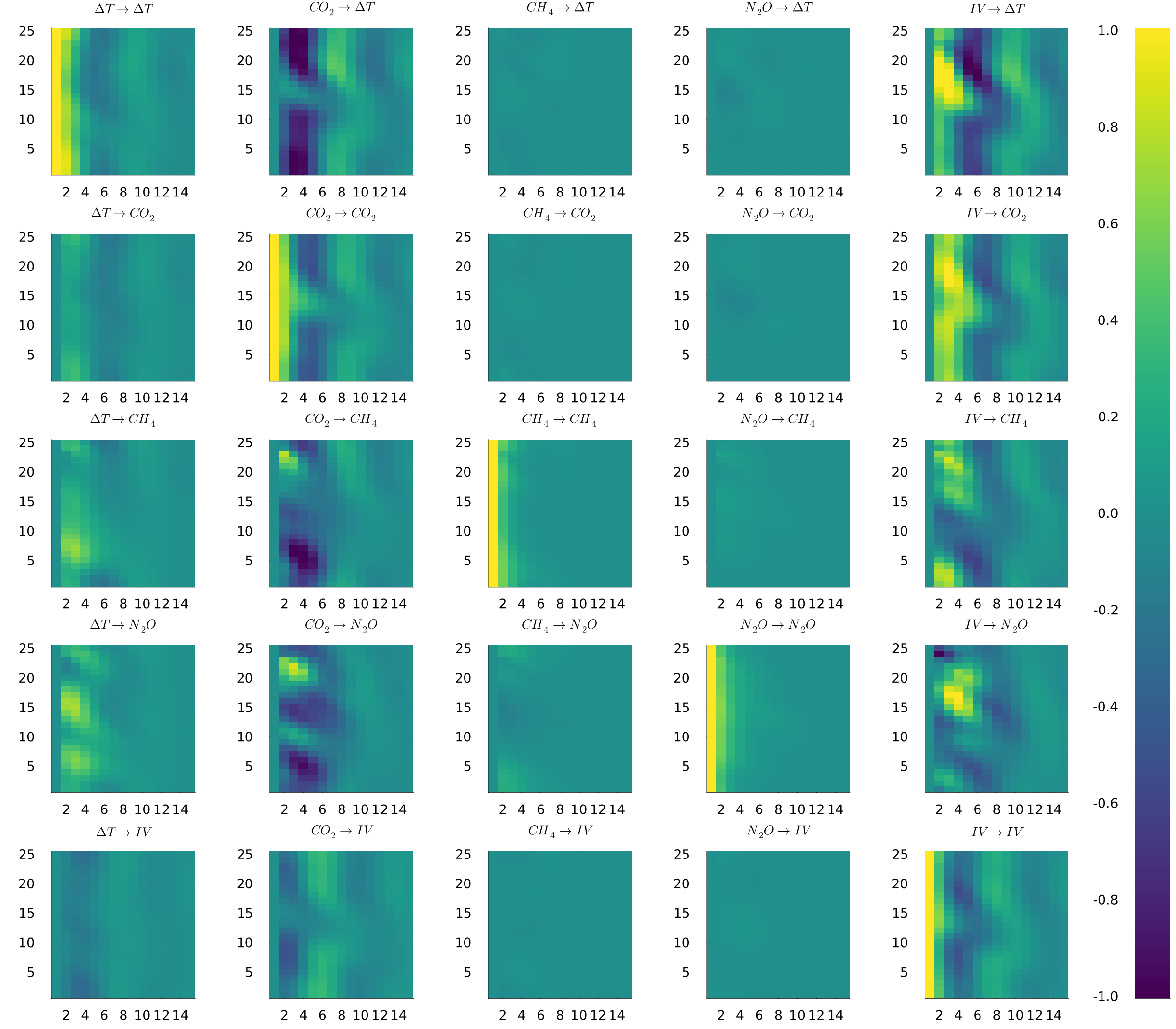} 
\caption{Matrix elements of the time dependent response function obtained from the model coefficients plotted as a function of time. The response function is given by Eq. \eqref{Response_function_model}.}
\label{fig:Response_function_model}
\end{figure*}

Knowledge of the estimated model coefficients enables us to construct the linear response function, $R(\tau;t)$, which delineates the causal
relationships among the time series under consideration. For the model
analyzed here, $R(\tau;t)$ can be expressed in terms of the time-dependent correlation (or persistence) matrix, $C(\tau;t)$, following
\citet{baldovin2022extracting}, as
\begin{equation}
    R(\tau;t) = C(\tau;t) C^{-1}(\tau;0).
    \label{Response_function_data}
\end{equation}
\textcolor{black}{Here, $\tau$ denotes the reference time, or equivalently
the phase within the periodic coefficient cycle, while $t$ denotes the
time lag over which the correlation and corresponding response are
evaluated.}

The elements of the time-periodic correlation matrix are defined by
\begin{equation}
    C_{ij}(\tau;t) =
    \frac{1}{M}
    \sum_{n=0}^{M-1}
    \eta_i(\tau+t+nT\Delta t)
    \eta_j(\tau+nT\Delta t),
    \label{Correlation_matrix}
\end{equation}
where $M$ and $T$ are specified in Sec. \ref{Stochastic_models_1D}.
\textcolor{black}{Since $T\Delta t=P$, the term $nT\Delta t=nP$ shifts the
reference time by an integer number of complete periods. Therefore, Eq.
(\ref{Correlation_matrix}) estimates the correlation between
$\eta_j$ at a given phase of the cycle and $\eta_i$ a time $t$ later by
averaging over $M$ realizations of the same phase. This phase-conditioned averaging is required because the non-autonomous model is periodically nonstationary: unlike a stationary correlation function, the correlation depends not only on the lag $t$ but also on the position $\tau$ within the coefficient cycle. It therefore retains the periodic modulation of the statistical dependence between the climate variables while improving the estimate through averaging over repeated cycles.  This underlies the memory-effect described by \citet{moon2017unified}.}

For the fitted linear nonautonomous model, the response matrix can also be obtained from the drift matrix $\boldsymbol K(t)$, whose elements are given by $K_{ii}(t)=a_i(t)-\sum_{j\neq i}b_{ij}(t)$ and $K_{ij}(t)=b_{ij}(t)$ governing the deterministic part of the dynamics. In general, the propagator of such a system is given by the time-ordered exponential
\begin{equation}
R(\tau;t)=\mathcal{T}\exp\!\Big[\int_{\tau}^{\tau+t}\boldsymbol{K}(t')\,dt'\Big],
   \label{Response_function_model_exact}
\end{equation}
where $\mathcal{T}$ denotes the time-ordering operator, which arranges matrix products in chronological order when the drift matrices at different times do not commute. In the present application, however, the autocorrelation functions predicted by the fitted model decay to near zero on timescales substantially shorter than the period of the cyclostationary coefficients. This indicates that the dominant response is confined to lags over which the temporal modulation of $\boldsymbol K(t)$ has only a limited cumulative effect. Under this short-memory approximation we therefore neglect the commutator corrections associated with explicit time ordering and approximate the propagator as
\begin{equation}
   R(\tau;t)\approx
   \exp\!\Big[\int_{\tau}^{\tau+t}\boldsymbol{K}(t')\,dt'\Big].
   \label{Response_function_model}
\end{equation}

The causal relationships between ice volume, temperature, carbon dioxide, methane, and nitrous oxide during the Late Pleistocene constitute a hierarchical and asymmetric network of interactions that operate across multiple timescales. In Figures \ref{fig:Response_function_data} and \ref{fig:Response_function_model}, we present the time evolution of the matrix elements of the response function, which are derived from the time series of these five variables and the estimated model coefficients, respectively. Despite the inherent noise present in Figure \ref{fig:Response_function_data}, which is a consequence of averaging over a limited sample size of $M = 34$ points [see Eq. \eqref{Response_function_data}], it agrees qualitatively with the findings shown in Figure \ref{fig:Response_function_model}.

\textcolor{black}{The two approaches exhibit broadly consistent temporal structures for the dominant response-function elements, including the occurrence and characteristic lag dependence of their principal variations. Quantitative differences remain, however, particularly in the amplitudes of the responses and in the precise locations and depths of their extrema. 
The data-based estimates of the correlation matrices are constructed from only $M=34$ realizations at each phase, making the resulting response functions sensitive to finite-sample variability. Moreover, the inversion of the correlation matrix in Eq. \eqref{Response_function_data} can amplify this sampling variability. In contrast, the coefficient-based response functions are considerably smoother because they are calculated directly from the inferred dynamical operator and therefore do not contain the same finite-sample fluctuations. Thus, the agreement in the dominant signs, relative response strengths, asymmetries, and characteristic temporal structure constitutes the most robust correspondence between the two estimates, whereas differences in their detailed amplitudes and extrema should be interpreted more cautiously.}


Our analysis indicates that atmospheric CO$_2$ exerts a more pronounced influence on ice volume (IV) than does temperature. This observation is reflected in the fact that the magnitude of the time averaged coupling coefficient associated with $\text{CO}_2 \rightarrow \text{IV}$ is larger than that associated with $\Delta\text{T} \rightarrow \text{IV}$. Additionally, we find that the causal relationship from CO$_2$ to $\Delta$T is stronger than that from $\Delta$T to CO$_2$, corroborating the findings of \citet{baldovin2022extracting} and Keyes \textit{et al.} \cite{keyes2023stochastic}. Therefore, CO$_2$ controls $\Delta$T and IV. Furthermore, our findings suggest that CH$_4$ and N$_2$O have negligible influence on IV, $\Delta$T, and CO$_2$, whereas, we observe that IV, $\Delta$T and CO$_2$ have a strong causal link to, and hence a strong influence on, CH$_4$ and N$_2$O. This analysis of response functions is consistent with the model interpretation discussed in Section \ref{Model_interpretation}.

An important insight is that ice volume provides an additional degree of freedom, whose coupling structure--rather than a difference in intrinsic relaxation timescales--enables oscillatory behavior in response functions. These oscillations were not captured in our earlier model \cite{keyes2023stochastic}, which did not include IV, whereas here they arise naturally and their oscillatory dynamics can be explained as follows. Over a time window short compared with the cyclostationary modulation, one may regard the coefficients as approximately constant and focus on the dominant subdynamics between $\Delta \text{T}, \mathrm{CO}_2$ and $\mathrm{IV}$.  For example, expanding the diffusive coupling terms in Eq.~\eqref{eqn:5D_model} for the two-variable case involving $\Delta \text{T}$ and $\mathrm{IV}$, the dynamics become
\begin{equation}
\frac{d}{dt}
\begin{pmatrix}
\eta_{\Delta T}\\[2pt]
\eta_{\mathrm{IV}}
\end{pmatrix}
=
\underbrace{
\begin{pmatrix}
a_{\Delta T}-b_{\Delta T,\mathrm{IV}} & b_{\Delta T,\mathrm{IV}}\\
b_{\mathrm{IV},\Delta T} & a_{\mathrm{IV}}-b_{\mathrm{IV},\Delta T}
\end{pmatrix}}_{\boldsymbol{K}}
\begin{pmatrix}
\eta_{\Delta T}\\[2pt]
\eta_{\mathrm{IV}}
\end{pmatrix}
+ \text{noise}.
\label{eq:K}
\end{equation}
Here, the drift matrix $\boldsymbol{K}$ has diagonal elements representing the net stability of each variable (intrinsic stability $a_i$ minus the outflow to the coupled partner due to $b_{ij}$), while the off diagonal elements $b_{ij}$ represent the direct coupling from variable $j$ to variable $i$. The eigenvalues of $\boldsymbol{K}$ are
\begin{equation}
\lambda_{\pm}
=
\frac{\mathrm{Tr}(\boldsymbol{K})}{2}
\pm
\sqrt{
\frac{\mathrm{Tr}(\boldsymbol{K})^2}{4}
-\det(\boldsymbol{K})
},
\label{eq:eigs}
\end{equation}
where $\mathrm{Tr}(\boldsymbol{K}) = (a_{\Delta \text{T}}-b_{\Delta \text{T},\mathrm{IV}}) + (a_{\mathrm{IV}}-b_{\mathrm{IV},\Delta \text{T}})$ is the trace and $\det(\boldsymbol{K}) = (a_{\Delta \text{T}}-b_{\Delta \text{T},\mathrm{IV}})(a_{\mathrm{IV}}-b_{\mathrm{IV},\Delta \text{T}}) - b_{\Delta \text{T},\mathrm{IV}}\,b_{\mathrm{IV},\Delta \text{T}}$ is the determinant.

A damped oscillatory mode (complex-conjugate $\lambda_{\pm}$ with negative real part) requires two conditions: (i) stability, $\mathrm{Tr}(\boldsymbol{K})<0$, and (ii) a negative discriminant, $\mathrm{Tr}(\boldsymbol{K})^2 < 4\det(\boldsymbol{K})$. When the diagonal stabilities are comparable, as is the case for $\Delta T$ and IV here, the second condition simplifies to approximately $b_{\Delta \text{T},\mathrm{IV}}\,b_{\mathrm{IV},\Delta \text{T}}<0$, so that the cross-couplings must have opposite signs. Similarly, we can show that $b_{\mathrm{CO}_2,\mathrm{IV}}\,b_{\mathrm{IV},\mathrm{CO}_2}<0$. This is precisely the structure inferred from the data: the couplings $\Delta \text{T}\to \mathrm{IV}$ and $\mathrm{CO}_2\to \mathrm{IV}$ are negative, so that temperature and $\mathrm{CO}_2$ destabilize---promote the growth of---ice volume, while the reverse couplings $\mathrm{IV}\to \Delta \text{T}$ and $\mathrm{IV}\to \mathrm{CO}_2$ are positive, so that ice volume stabilizes--suppresses the growth of--temperature and $\mathrm{CO}_2$. The oscillation period is determined by $2\pi/|\mathrm{Im}(\lambda_{\pm})|$, which depends on the interplay between the net stabilities and the coupling magnitudes through the determinant. In contrast, the $\Delta \text{T} - \mathrm{CO}_2$ couplings are predominantly of the same sign (both positive), so their two-dimensional sub-block yields purely relaxational (non-oscillatory) responses.

Physically, opposite-sign cross-couplings, and hence oscillatory behaviors, provide a consistent interpretation once one recognizes the role of strong ice-volume feedback on both $\Delta$T and CO$_2$ through its influence on albedo, surface topography, ocean circulation, land-atmosphere and ocean–atmosphere carbon exchange. For example, IV controls how stored ocean heat and carbon anomalies are communicated to the atmosphere. Because ocean heat content and the surface carbon system adjust through mixed-layer and circulation processes, the associated feedback is intrinsically non-instantaneous, providing the phase lag needed for oscillatory behavior. In Figures \ref{fig:Response_function_data} and \ref{fig:Response_function_model}, we clearly observe that ice volume feedbacks lag both $\Delta$T and CO$_2$, consistent with earlier studies by \citet{bintanja2008north} and \citet{shakun2015800}. \textcolor{black}{The earlier four-variable analysis cannot capture these findings. Hence, the principal novelty is not simply the estimation of additional IV-related coefficients, but the emergence of a qualitatively different response structure that results from the explicit incorporation of cryosphere--climate feedbacks.} In summary, these causal relationships support the perspective that ice volume functions as a slow integrator and regulator of climate forcing, while temperature and greenhouse gases serve as faster amplifying feedback mechanisms that shape the amplitude, persistence, and timing of Late Pleistocene glacial cycles.

\section{Conclusion} \label{Conclusions}

Over the past 800 kyr, Earth’s paleoclimate has exhibited periodic yet intrinsically noisy glacial–interglacial cycles, with a characteristic period of approximately 100 kyr, clearly reflected in proxy time series of ice volume, temperature, carbon dioxide, methane, and nitrous oxide derived from marine $\delta^{18}$O records and EPICA ice core record. We applied multifractal time-weighted detrended fluctuation analysis (MFTWDFA) to these time series to quantify their scale-dependent stochastic structure, identify the dominant colored-noise regimes, and determine crossover times between distinct dynamical behaviors with greater precision than what conventional spectral slope analysis allows. This analysis provided a theoretically grounded basis for representing paleoclimate time series using non-autonomous stochastic models.  

Building on this foundation, we formulated both single-variable and five-variable stochastic dynamical models that explicitly account for the time-dependent deterministic and stochastic components of the climate records. After identifying the timescale-separated colored noise regimes in the data, we computed deterministic stability, noise amplitude, and inter-variable coupling coefficients using non-autonomous Ornstein-Uhlenbeck models. These coupling coefficients modulate both the magnitude and directionality of interactions among the variables, thereby elucidating how stabilizing and destabilizing feedbacks operate across multiple timescales and providing insight into the dynamical structure underlying Late Pleistocene climate variability.

A key qualitative difference between the present framework and our earlier stochastic paleoclimate model \cite{keyes2023stochastic} is the explicit inclusion of ice volume (IV) as a dynamical variable. \textcolor{black}{The present five-variable model provides both a robustness test of the interaction structure reported earlier and a means of identifying changes associated with the explicit representation of cryospheric dynamics.} From the estimated coefficients, the net stability of IV is comparable to that of $\Delta \text{T}$ and $\mathrm{CO}_2$; all three variables have similar (relatively weak) mean-reversion rates. In contrast, $\mathrm{CH}_4$ and $\mathrm{N}_2\mathrm{O}$ are markedly more stable (more strongly mean-reverting) and therefore relax on substantially shorter timescales relative to the dominant climate variability. Consequently, $\mathrm{CH}_4$ and $\mathrm{N}_2\mathrm{O}$ contribute primarily to short-lived transients and do not control the long-lag structure of the response. We found that both temperature and carbon dioxide exert destabilizing influences on ice volume by regulating ice-sheet mass balance through thermal forcing and greenhouse radiative effects. Ice volume, in turn, exerts stabilizing influences on both temperature and carbon dioxide through the ice-albedo feedback, the effects of elevation, and ocean–carbon cycle processes.

We observed oscillatory dynamics in the response functions and concluded that this behavior stems from the opposite-sign cross-couplings between IV and $\Delta \text{T}$, and IV and $\mathrm{CO}_2$. Our response-function analysis further demonstrates a strong causal relation from CO$_2$ to ice volume and temperature, thereby highlighting the dominant role of atmospheric CO$_2$ as greenhouse forcing of glacial–interglacial transitions. Methane and nitrous oxide respond sensitively to ice-driven boundary conditions, temperature, and ocean circulation, although they exert a weak direct control on ice volume, temperature, and carbon dioxide. Our model also captures the observed lagged feedbacks of ice-volume on temperature and CO$_2$, consistent with its role as a slow integrator of climate forcing.

Our approach provides a systematic stochastic modeling framework for quantifying glacial–interglacial climate dynamics.
Future developments should incorporate nonlinear intervariable coupling, and multiplicative and correlated noise, to better reflect the inherently nonlinear nature of the climate system. A natural extension of the framework is to include additional paleoclimate variables, such as ocean circulation proxies and dust, which may further clarify the pathways through which climate feedbacks operate. Although such extensions will increase model dimensionality, the resulting formulations will remain orders of magnitude simpler than comprehensive global climate models. By reproducing key statistical and dynamical properties of paleoclimate records, this approach offers a promising method for constraining, testing, and interpreting complex climate models across a wide range of past climate states, and for guiding future investigations of Earth system sensitivity and predictability.

\section*{ACKNOWLEDGMENTS} P.P. and J.S.W.  gratefully acknowledge support from the Swedish Research Council under Grant No.638-2013-9243. Nordita is partially supported by Nordforsk.

\section*{AUTHOR DECLARATIONS}
The authors have no conflicts to disclose.

\section*{DATA AVAILABILITY}
The data that support the findings of this study are available from the corresponding author upon reasonable request.

\appendix

\color{black}
\section{Multi-scale analysis of Figure \ref{fig:mftwdfa}{\label{sec:app_A}}}

Figure \ref{fig:mftwdfa} can be interpreted by separating the dynamics between timescales much shorter than, comparable to, and much longer than the modulation period of 
\begin{equation}
P=23.5~\mathrm{kyr}.
\nonumber
\end{equation}
\subsection{Short timescales: $s\ll P$}

On these short timescales, the coefficients of the non-autonomous Ornstein--Uhlenbeck process
\begin{equation}
dX=a(t)X\,dt+N(t)\,dW_t
\end{equation}
change very little throughout a window of duration $s$. Therefore, locally we can approximate
\begin{equation}
a(t)\simeq a(t_0),
\qquad
N(t)\simeq N(t_0),
\end{equation}
so that the dynamics can be treated as those of an autonomous OU process. The red-noise response to white noise forcing is reflected in the power spectrum 
\begin{equation}
S(f)\sim f^{-\beta}, \qquad \text{where} \qquad \beta =2.
\end{equation}
Thus, the Hurst exponent $h(q=2)$ is
\begin{equation}
h(2)=\frac{1+\beta}{2} = \frac{3}{2}.
\end{equation}
This explains the approximately red-noise slopes observed at short timescales in both Figs. \ref{fig:mftwdfa}(a) and (b).

\subsection{Intermediate timescales: $s\sim P$}

On these timescales, the coefficients vary appreciably during the observation window and thus the dynamics are unambiguously non-autonomous. Therefore, the fluctuation function is not expected to follow a single power law, and it is more appropriate to consider the scale-dependent slope viz.
\begin{equation}
h_{\mathrm{eff}}(s)
=
\frac{d\log F_2(s)}{d\log s}.
\end{equation}
Thus, the bending of the curves in the vicinity of the $P=23.5$ kyr timescale corresponds to a regime in which the effective Hurst exponent varies continuously with scale.

\subsection{Long timescales: $s\gg P$}

At long timescales, the flattening observed in Figs. \ref{fig:mftwdfa}(a) and (b) has two different origins. 

Consider first the case of Fig. \ref{fig:mftwdfa}(a).  Before filtering, the original records contain a strong approximately $100$ kyr glacial-cycle component. Consider, for simplicity,
\begin{equation}
X(t)=A\cos\left(\frac{2\pi t}{P_g}\right),
\qquad
P_g\simeq100~\mathrm{kyr}.
\end{equation}
The profile used in MFTWDFA is the integral of the signal,
\begin{equation}
Y(t)
=
\int_0^t X(u)\,du
=
\frac{AP_g}{2\pi}
\sin\left(\frac{2\pi t}{P_g}\right).
\end{equation}
Because this profile is bounded,
\begin{equation}
|Y(t)|\leq \frac{AP_g}{2\pi},
\end{equation}
once the observation window contains one or several complete glacial cycles, positive and negative contributions increasingly cancel. The fluctuation function no longer grows as $\sqrt{s}$ but instead approaches a constant:
\begin{equation}
F_2(s)\rightarrow \mathrm{constant},
\label{eq:F2const}
\qquad
h_{\mathrm{eff}}(s)\rightarrow0.
\end{equation}
Since the dominant oscillation has a period of order $100$ kyr, this explains why the flattening in Fig. \ref{fig:mftwdfa}(a) occurs only at comparatively large values of $s$.

Now consider the case of Fig. \ref{fig:mftwdfa}(b).  The fluctuation signal is obtained by subtracting a Gaussian-smoothed version of the original signal,
\begin{equation}
X_{\mathrm{hp}}(t)
=
X(t)-G_\sigma*X(t).
\end{equation}
For a Gaussian kernel,
\begin{equation}
G_\sigma(t)
=
\frac{1}{\sqrt{2\pi}\sigma}
\exp\left(-\frac{t^2}{2\sigma^2}\right),
\end{equation}
whose Fourier transform is
\begin{equation}
\widehat G_\sigma(f)
=
\exp\left[-\frac{(2\pi f\sigma)^2}{2}\right].
\end{equation}
Therefore,
\begin{equation}
\widehat X_{\mathrm{hp}}(f)
=
\left[
1-
\exp\left(-\frac{(2\pi f\sigma)^2}{2}\right)
\right]\widehat X(f),
\end{equation}
and the corresponding high-pass transfer function is
\begin{equation}
H_{\mathrm{hp}}(f)
=
1-
\exp\left[-\frac{(2\pi f\sigma)^2}{2}\right].
\end{equation}

The filtered power spectrum is given by
\begin{equation}
S_{\mathrm{hp}}(f)
=
|H_{\mathrm{hp}}(f)|^2 S(f), 
\end{equation}
so that at low frequency,
\begin{equation}
2\pi f\sigma\ll1,
\end{equation}
and thus
\begin{equation}
\exp\left[-\frac{(2\pi f\sigma)^2}{2}\right]
\simeq
1-\frac{(2\pi f\sigma)^2}{2}.
\end{equation}
Therefore, 
\begin{equation}
H_{\mathrm{hp}}(f)
\simeq
\frac{(2\pi f\sigma)^2}{2},
\end{equation}
and hence
\begin{equation}
S_{\mathrm{hp}}(f)
\propto
f^4 S(f),
\qquad f\rightarrow0.
\end{equation}
Thus the high-pass filtering strongly suppresses the low-frequency power. As a result, the integrated MFTWDFA profile does not continue accumulating variance at large $s$, and for this reason 
\begin{equation}
F_2(s)\rightarrow \mathrm{constant},
\qquad
h_{\mathrm{eff}}(s)\rightarrow0.
\end{equation}

Therefore, the two figures flatten for different reasons. In Fig. \ref{fig:mftwdfa}(a), the flattening is principally associated with the bounded, approximately $100$ kyr glacial-cycle variability already present in the original records. In Fig. \ref{fig:mftwdfa}(b), the flattening is instead produced by the explicit suppression of low-frequency power introduced by the high-pass filter. Because these two mechanisms act on different characteristic timescales, the flattening of the curves in panel (b) occurs at smaller $s$ than in panel (a), thereby producing the observed time shift between them.

\color{black}
\section{Weak coupling example{\label{sec:app_B}}}

\begin{figure*}
\centering
\begin{subfigure}[b]{1.0\textwidth}
    \includegraphics[width=\linewidth]{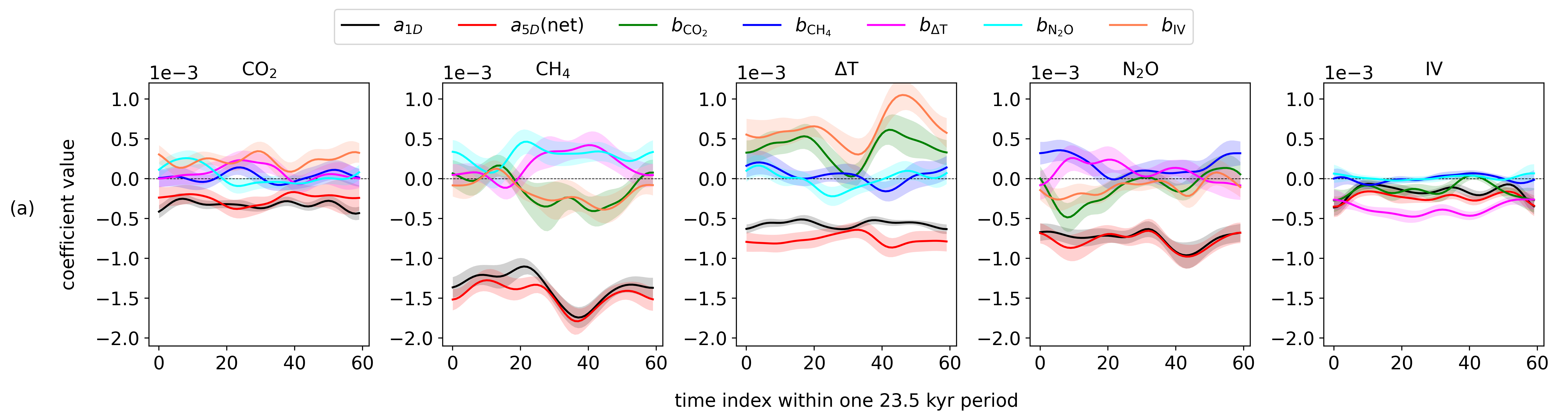}
\end{subfigure}
\hfill
\begin{subfigure}[b]{1.0\textwidth}
    \includegraphics[width=\linewidth]{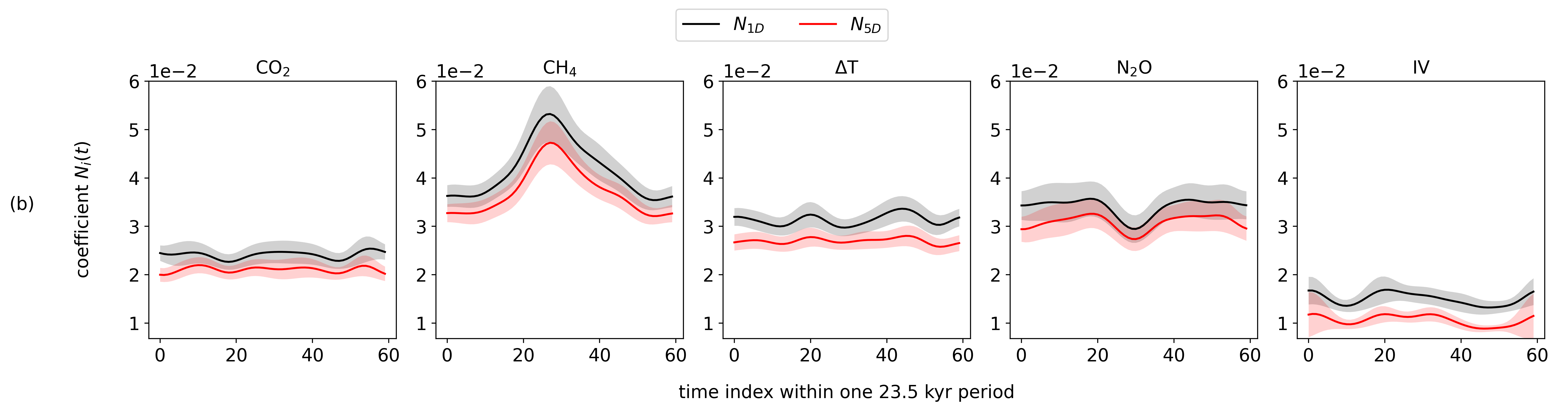}
\end{subfigure}
\caption{Same as Fig. \ref{fig:model_coefficients}, but computed using the gap-filled N$_2$O time series and a higher temporal resolution of $60$ data points per period for all variables.}
\label{fig:model_coefficients_gap_filled_N2O_data}
\end{figure*}

Let $\eta_i$ denote either $\text{N}_2\text{O}$ or $\text{CH}_4$ and $\eta_2$ denote the ice volume, with the time averaged coefficients written using overbars giving
\begin{align}
   \frac{d \eta_i}{d t}&=\left(\overline{a_i}-\overline{b_{i 2}}\right) \eta_i(t)+\overline{b_{i 2}} \eta_2(t)+N_i(t) \xi_i(t),
\label{eq:ghg} \\
\frac{d \eta_2}{d t}&=\left(\overline{a_2}-\overline{b_{2 i}}\right) \eta_2(t)+\overline{b_{2 i}} \eta_i(t)+N_2(t) \xi_2(t) \nonumber\\
&\approx - |\overline{a_2}| \eta_2(t)+ N_2(t) \xi_2(t), 
\label{eq:iv} 
\end{align}
with approximate values from Fig.~\ref{fig:model_coefficients} of $\overline{a_i} = -7$, $\overline{a_2} = -2$, $\overline{b_{i 2}} = -1$ and $\overline{b_{2 i}}$=0, so that the ice volume dynamics is an Ornstein-Uhlenbeck process, the solution to which is 
\begin{equation}
\eta_2(t)=\eta_2(0) e^{-|\overline{a_2}| t}+e^{-|\overline{a_2}| t} \int_0^t e^{|\overline{a_2}| t^{\prime}} N_2(t^\prime) \xi_2(t^\prime) d t^{\prime}.
\label{eq:ivsln}
\end{equation}
We either consider the deterministic dynamics or the ensemble average viz.,
\begin{equation}
   \langle \eta_2(t)\rangle=\eta_2(0) e^{-|\overline{a_2}| t}\equiv e^{-|\overline{a_2}| t}.
\end{equation}
Using   $\langle \eta_2(t)\rangle$ in Eq. \eqref{eq:ghg} with $\eta_i(0)=1$ we obtain 
\begin{align}
       \eta_i(t)&=e^{-\left(\left|\overline{a_i}\right|-\left|\overline{b_{i 2}}\right|\right) t}\left[\dfrac{\left|\overline{a_i}\right|-\left|\overline{a_2}\right|-\left|\overline{b_{i 2}}\right| \nonumber e^{\left(\left|\overline{a_i}\right|-\left|\overline{a_2}\right|-\left|\overline{b_{i 2}}\right|\right) t}}{\left|\overline{a_i}\right|-\left|\overline{a_2}\right|-\left|\overline{b_{i 2}}\right|}\right] \nonumber \\ &+ 
   e^{-\left(\left|\overline{a_i}\right|-\left|\overline{b_{i 2}}\right|\right) t} \int_0^t e^{\left(\left|\overline{a_i}\right|-\left|\overline{b_{i 2}}\right|\right) t^{\prime}} N_i(t^\prime) \xi_i(t^\prime)
d t^{\prime},
\end{align}
so that 
\begin{equation}
   \langle \eta_i(t)\rangle =e^{-\left(\left|\overline{a_i}\right|-\left|\overline{b_{i 2}}\right|\right) t}\left[\frac{\left|\overline{a_i}\right|-\left|\overline{a_2}\right|-\left|\overline{b_{i 2}}\right| e^{\left(\left|\overline{a_i}\right|-\left|\overline{a_2}\right|-\left|\overline{b_{i 2}}\right|\right) t}}{\left|\overline{a_i}\right|-\left|\overline{a_2}\right|-\left|\overline{b_{i 2}}\right|}\right].
\end{equation}
As the ice volume coupling strength increases, and $|\overline{b_{i 2}}|\rightarrow |\overline{a_i}|$, which is a limit not realized (see Fig.~\ref{fig:model_coefficients}a), the decay of $\langle \eta_i(t)\rangle$ from the initial condition is controlled by the ice volume, $|\overline{a_2}|$, but is more rapid at short times than $\langle \eta_2(t)\rangle$, and is 
\begin{equation}
\left\langle\eta_i(t)\right\rangle=1+\frac{\left|\overline{a_i}\right|}{\left|\overline{a_2}\right|}\left(e^{-\left|\overline{a_2}\right| t}-1\right),
\end{equation}
rapidly becoming negative at time
\begin{equation}
t^* = \frac{1}{\left|\overline{a_2}\right|}\left[\log(\left|\overline{a_i}\right|) - \log(\left|\overline{a_i}\right|-\left|\overline{a_2}\right|)\right], \label{time_t_star}
\end{equation}
with $t^*$=1.68 kyr for the values used here, whereas $\langle \eta_2(t)\rangle$ shows simple exponential decay and thus never changes sign.  Therefore, for the most extreme--and never realized--ice volume coupling constant the stochastic process of the ice volume is mean reverting to the origin and that for either greenhouse gas reverts to negative values, with the deterministic decay of both controlled by that of the ice volume.  

Note that Eq \eqref{eq:ivsln} is trivial for $\eta_2(t=0)=0$, and if we either assume $N_2(t)$ is constant in time or an arbitrary function of time we have the same result because $\xi_2(t)$ is a Gaussian White Noise process. Therefore the integral is also a stochastic process. For simplicity assume $N_2(t)$ constant in time, so does not contribute to the integral
\begin{equation}
Y(t)= \int_0^t e^{|\overline{a_2}| t^{\prime}} \xi_2(t^\prime) d t^{\prime},
\end{equation}
where $\xi_2(t)=\dfrac{d W(t)}{d t}$ is the derivative of a Wiener process $W(t)$, such that
\begin{equation}
Y(t)= \int_0^t e^{|\overline{a_2}| t^{\prime}} dW(t^{\prime}),
\end{equation}
which is a stochastic (It$\hat{\text{o}}$) integral, and thus is another stochastic process. Specifically,
$Y(t)$ is normally distributed with mean zero and variance 
\begin{equation}
\operatorname{Var}(Y(t))=\int_0^t e^{2|\overline{a_2}| t^{\prime}} d t^{\prime}=\frac{\exp (2 |\overline{a_2}|   t)-1}{2 |\overline{a_2}|}.
\end{equation}
Therefore, $Y(t=0)$=0=$\operatorname{Var}(Y(t=0))$.

\section{Model coefficients using gap-filled N$_2$O data {\label{sec:app_C}}}

Recently, \citet{salehnia2025continuous} reconstructed a physically consistent and continuous record of N$_2$O from an incomplete record over the past 800 kyr using a machine learning technique called Gaussian Process Regression with a Rational Quadratic kernel (GPR–RQ). They applied this method to EPICA ice-core data on CO$_2$, CH$_4$, and N$_2$O, successfully addressing gaps in the N$_2$O record caused by contamination and sparse measurements during glacial periods. Here, we examine the implications of using this gap-filled N$_2$O time series on the interpretation of our model.

The data used to generate the results discussed in the main body of the paper contained the following number of data points: 8001 for CO$_2$, temperature, and ice volume; 2103 for CH$_4$; and 912 for N$_2$O. The gap-filled N$_2$O time series contains 7915 data points \citep{salehnia2025continuous}. To model the $23.5$-kyr frequency, which corresponds to approximately 34 cycles within the $800$-kyr interval, we employ two distinct discretization strategies. The first approach, consistent with our analysis in the main body of the paper, considers $25$ data points per period. Application of this strategy to the gap-filled N$_2$O data yields model coefficients that are nearly indistinguishable from those obtained using the original EPICA record for N$_2$O. In our second approach, which accounts for the lower resolution of the CH$_4$ time series, we use 60 data points per period across all variables. Although this higher-resolution sampling leads to a slight increase in the magnitudes of the time-averaged coefficients, the overall structure of the coupling and stability coefficients remains unchanged (See Fig. \ref{fig:model_coefficients_gap_filled_N2O_data} cf. Fig. \ref{fig:model_coefficients}).

Most importantly, the analysis of the time-averaged model coefficients obtained from the second approach further reinforces the validity of the weak coupling argument presented in Appendix \ref{sec:app_B}. The relevant coefficients in Eqs. \eqref{eq:ghg} and \eqref{eq:iv} for this case in Fig. \ref{fig:model_coefficients_gap_filled_N2O_data} are as follows: $\overline{a_i} = -8$, $\overline{a_2} = -2$, $\overline{b_{i 2}} = -1$ and $\overline{b_{2 i}}=0$. Using these values, the time $t^*$ from Eq. \eqref{time_t_star} is $1.438$ kyr. This confirms the robustness of our stochastic framework to variations in temporal discretization and data resolution, and thus the resulting coefficients capture the essential dynamics of the system. 

\providecommand{\noopsort}[1]{}\providecommand{\singleletter}[1]{#1}%

\end{document}